\input amstex
\input amsppt.sty
\input epsf.sty

\TagsOnRight
\NoBlackBoxes
\NoRunningHeads
\magnification 1200

\hoffset=0.5cm
\voffset=-1cm

\define\x{\frak X}
\define\y{\frak Y}
\define\z{\frak Z}
\define\ze{\zeta}
\define\C{\Bbb C}

\define\Z{\Bbb Z}
\define\Ma{\operatorname{Mat}}
\define\r{\underset{\zeta=x}\to{\operatorname{Res}}\,}
\define\F{\Cal F}
\define\G{\Cal G}
\define\tht{\thetag}

\define\Ga{\Gamma}
\define\si{\sigma}
\define\q{q_k^{(z,z',\xi)}}
\topmatter

\author Alexei Borodin 
\endauthor

\title Discrete gap probabilities and discrete Painlev\'e equations
\endtitle

\abstract We prove that Fredholm determinants of the form $\det(1-K_s)$, where $K_s$ is the restriction of either the discrete Bessel kernel or the discrete ${}_2F_1$ kernel to $\{s,s+1,\dots\}$, can be expressed through solutions of discrete Painlev\'e II and V equations, respectively.

These Fredholm determinants can also be viewed as distribution functions of the first part of the random partitions distributed according to a poissonized Plancherel measure and a z-measure, or as normalized Toeplitz determinants with symbols $e^{\eta(\ze+\ze^{-1})}$ and $(1+\sqrt{\xi}\zeta)^z(1+\sqrt{\xi}/\ze)^{z'}$.

The proofs are based on a general formalism involving discrete integrable operators and discrete Riemann--Hilbert problem. A continuous version of the formalism has been worked out in \cite{BD}.

\endabstract

\toc
\widestnumber\head{??}
\head {} Introduction \endhead
\head 1. Integrable operators and discrete Riemann--Hilbert problems\endhead
\head 2. Simpler DRHP for a class of integrable operators \endhead
\head 3. Discrete Bessel kernel and dPII
\endhead 
\head 4. Fredholm determinant and dPII \endhead 
\head 5. Initial conditions for dPII \endhead
\head 6. Discrete ${}_2F_1$ kernel and dPV \endhead
\head 7. Fredholm determinant and dPV\endhead
\head 8. Initial conditions for dPV \endhead
\head 9. Degeneration to continuous PII and PV \endhead
\head {} References \endhead
\head {} Picture \endhead
\endtoc

\endtopmatter

\document 
\head Introduction
\endhead

In recent years we have witnessed a discovery and an intensive study of a class of discrete probabilistic models which in appropriate limits converge, in one way or another, to well--known models of Random Matrix Theory (RMT, for short). The sources of new models are quite diverse, they include Combinatorics, Representation Theory, Percolation Theory, Growth Processes, tiling problems and others. 

One quantity of interest in RMT is the gap probability --- the probability of having no particles--eigenvalues in a given interval. In particular, the level spacing distribution and the distribution of the most right or left particle are easily expressible in terms of  gap probabilities. Naturally, gap probabilities also arise as limits of relevant quantities of the discrete models mentioned above. In many cases these quantities can be viewed as gap probabilities for certain discrete random point processes. 

It is well known that in random matrix models gap probabilities can often be expressed in terms of a solution of a 2nd order nonlinear ordinary differential equation which, quite remarkably, happens to be one of the six Painlev\'e equations, see, e.g., \cite{TW2}. 

In this paper we show that the counterparts of gap probabilities in two important discrete models can be expressed through solutions of discrete analogs of Painlev\'e II and Painlev\'e V equations. We also develop a general formalism which allows us to handle these 2 examples, and which we expect to work in other models as well.

Let us describe our results.

Let $S_n$ be the symmetric group of degree $n$. Denote by $l_n(\sigma)$ the length of the longest increasing subsequence of a permutation $\sigma\in S_n$. Set
$$
p_k^{n}=\frac 1{n!}\operatorname{Card}\{\sigma\in S_n\,|\,l_n(\sigma)\le k\}, \qquad p_k^{(\eta)}=e^{-\eta^2}\sum_{n=0}^\infty \frac{\eta^{2n}}{n!}\,p_k^n.
\tag 0.1
$$
Here $\eta$ is a complex parameter. 

There are many other ways to define $p_k^{(\eta)}$. For example, thanks to a result of \cite{G}, it can be defined using a Toeplitz determinant
$$
p_k^{(\eta)}=e^{-\eta^2}\det[f_{i-j}]_{i,j=1}^k\,,\qquad \sum_{m=-\infty}^{+\infty} f_m\zeta^m=e^{\eta(\zeta+\zeta^{-1})}.
\tag 0.2
$$

A representation theoretic definition (which can be easily obtained using  the Robinson--Schensted correspondence) has the form
$$
p_k^{(\eta)}=e^{-\eta^2}\sum_{\lambda_1\le k} {\left(\frac{\dim\lambda}{|\lambda|!}\,\eta^{|\lambda|}\right)}^2
\tag 0.3
$$
where the summation is taken over all partitions $\lambda=(\lambda_1\ge\dots\ge \lambda_l>0)$ such that $\lambda_1\le k$, $|\lambda|=\lambda_1+\dots+\lambda_l$ is the size of the partition, and $\dim\lambda$ is the dimension of the irreducible representation of $S_{|\lambda|}$ corresponding to $\lambda$. 

A Fredholm determinant representation of $p_k^{(\eta)}$ relevant for us will be given in \S3 below.
\proclaim{Theorem 1} Let $\{x_n\}_{n=-1}^\infty$ be the sequence defined by $x_{0}=-1$, $x_1=f_1/f_0$ with $f_i$'s as in \tht{0.2}, and 
$$
x_{n+1}+x_{n-1}=\frac{nx_n}{\eta(x_n^2-1)}\,, \quad n\ge 0.
\tag 0.4
$$
Then for any $k\ge 1$ and generic $\eta$ we have {\rm (}dropping the superscript $(\eta)${\rm )}
$$
\frac{p_{k+1}p_{k-1}}{p_{k}^2}=1-x_k^2.
$$
\endproclaim
This result was also proved independently by J.~Baik \cite{Bai}. One more proof based on \cite{AvM} was found by M.~Adler and P.~van Moerbeke. The same equation for a related quantity was derived by C.~Tracy and H.~Widom in \cite{TW3}.

The relation \tht{0.4} is a special case of the {\it discrete PII equation}, see, e.g., \cite{GNR}.

Theorem 1 is a highly efficient tool for computing $p_k^{(\eta)}$ numerically. Indeed, the Toeplitz determinant interpretation gives the initial conditions
$p_0=e^{-\eta^2}$, $p_1=e^{-\eta^2}f_0$, and then Theorem 1 implies $p_{k+1}=(1-x_k^2)\,p_k^2/p_{k-1}$ for $k\ge 1$. Needless to say, this computational scheme is much faster than computing the Toeplitz determinants.

The celebrated result of \cite{BDJ1} claims that if we assume $\eta>0$ and let $\eta\to +\infty$, then $p_{2\eta+t\eta^{1/6}}^{(\eta)}$ converges to a smooth function $F_2(t)$ (also known as Tracy--Widom distribution in RMT \cite{TW1}) which can be expressed through a solution of the continuous PII equation. Given the existence of the limit, Theorem 1 implies that $(\ln F_2(t))''=-y^2(t)$ where $y(t)$ solves $y''=ty+2y^3$, which also follows from \cite{BDJ1}.

To state our second result we introduce the quantity $\q$ via the Toeplitz determinant
$$
\q=(1-\xi)^{zz'}\det[g_{i-j}]_{i,j=1}^k,\quad \sum_{m=-\infty}^{+\infty} g_m\zeta^m=(1+\sqrt{\xi}\zeta)^z(1+\sqrt{\xi}/\zeta)^{z'}.
\tag 0.5
$$
It is not hard to show that for any $k=1,2,\dots$, $\q$ extends to an analytic function in $(z,z',\xi)\in\C\times\C\times(\C\setminus [1,+\infty))$.

A representation theoretic definition of $\q$ says that if $z'=\bar{z}$ and $\xi\in(0,1)$ then $\q$ is the distribution function of the first row of the random Young diagram distributed according to a {\it z-measure}, see \cite{BO2}, \cite{BO3}. Z-measures are closely related to the generalized representations of the infinite symmetric group \cite{KOV}.
This interpretation of $\q$ leads to the identity, cf. \tht{0.3},
$$
\q=(1-\xi)^{zz'}\sum_{\lambda_1\le k}{\prod_{(i,j)\in D(\lambda)}(j-i+z)(j-i+z')}{\left(\frac{\dim\lambda}{|\lambda|!}
\right)}^2\xi^{|\lambda|}.
\tag 0.6
$$

Note that both \tht{0.5} and \tht{0.6} imply that 
$$
\q\to p_k^{(\eta)}\  \text{  as  }\  \xi\to 0,\ z,z'\to\infty,\ \xi zz'\to\eta^2.
$$

For positive integral values of $z,z'$, $\q$ also admits a longest increasing subsequence interpretation \cite{BO3, \S2}. It can also be viewed as the first passage time in an oriented percolation model, see \cite{J1} and also \cite{Bai}. For integral $z,z'$ of different signs, $\q$ gives a height distribution in a growth model called {\it digital boiling}, see \cite{GTW}. 

A Fredholm determinant representation of $\q$ which will be important for us, is given in \S6 below.

In what follows we denote by $F(a,b;c;u)$ the Gauss hypergeometric function.
\proclaim{Theorem 2} Let $\{x_n\}_{n=0}^\infty$ and $\{y_n\}_{n=0}^\infty$ be the sequences defined by the initial conditions
$$
\gather
x_{0}=-\frac{F(-z+1,\,-z';\,1;\,\xi)}{z'\xi\, F(-z+1,\,-z'+1;\,2;\,\xi)}\,,\\
y_0=\frac
{z'F(-z,\,-z'-1;\,1;\,\xi)\,F(-z+1,\,-z'+1;\,2;\,\xi)}
{F(-z,\,-z';\,1;\,\xi)\,F(-z+1,\,-z';\,2;\,\xi)}\,,
\endgather
$$
and the recurrence relations
$$
\gathered
x_{n+1}=\frac{\left(y_n-\left(z+z'+n+1\right)\right)
\left(y_n-\left(z'+n+1\right)\right)}
{\xi x_n y_n\left(y_n-z'\right)}\,,\\
y_{n+1}=-y_n+\frac{z+n+1}{1-x_{n+1}}+\frac{z'+n+2}
{1-\xi x_{n+1}}+z'.
\endgathered
\tag 0.7
$$
Then for any $k\ge 0$ and generic $(z,z',\xi)$ we have {\rm (}dropping the superscript $(z,z',\xi)${\rm )}
$$
\gathered
{\left(\frac{q_{k+1}}{q_k}-\frac{q_{k+2}}{q_{k+1}}\right)}
{\left(\frac{q_{k+2}}{q_{k+1}}-\frac{q_{k+3}}{q_{k+2}}\right)}^{-1}=
\frac{\left((1-\xi x_{k})(y_{k}-z')-(z+k+1)\right)}
{\left((1-\xi x_{k+1})(y_{k+1}-z')-(z+k+2)\right)}\\ \times \frac{(z+k+2)(z'+k+2)\left(y_k-(z'+k+1)\right)}{(1-\xi x_{k+1})(1-\xi x_k)\,
x_ky_k\left(y_k-z'\right)^2}\,.
\endgathered
\tag 0.8
$$
\endproclaim

Similar to Theorem 1, this theorem can be used for numerical evaluation of $\q$. Indeed, the initial conditions $q_0,q_1,q_2$ can be read off the Toeplitz determinant representation, and then \tht{0.8} provides a recurrence for computing $q_k$ for $k\ge 3$. The plot of the `probability density' $q^{(z,z',\xi)}_{k+1}-\q$ for certain specific values of $(z,z',\xi)$, computed using Theorem 2, can be found on the last page of the paper.

It is known that if $z'=\bar{z}$, $\xi\in(0,1)$, and $\xi\to 1$, then $q^{(z,z',\xi)}_{t/(1-\xi)}$ converges to a smooth function $G^{(z)}(t)$ which is the $\tau$-function of the Painlev\'e V equation, see \cite{BO2}, \cite{BD, \S8}. If we assume that $x_{t/(1-\xi)}$ has a smooth limit $x(t)$ then \tht{0.7} implies
$$
\gathered
x''(t)=\left(\frac 1{2x(t)}+\frac 1{x(t)-1}\right)(x'(t))^2-\frac{x'(t)}t+\frac{(z-z'-1)x(t)}t\\+\frac{(x(t)-1)^2}{2t}\left((z')^2x(t)-{z^2}/x(t)\right)-\frac 12\,\frac{x(t)(x(t)+1)}{x(t)-1}\,,
\endgathered
\tag 0.9
$$
which is a special case of the PV equation. Then the limit of \tht{0.8} provides an algebraic expression for $(\ln(\ln G^{(z)}(t))'')'$ in terms of $x(t)$ and $x'(t)$. In fact, the results of \cite{BD} imply that already $(\ln(G^{(z)})'$ can be algebraically expressed in terms of $x$ and $x'$ for a certain solution $x(t)$ of \tht{0.9}. It remains unknown whether there exists a discrete analog of this result. 

At present there exist several approaches to discretizing the Painlev\'e equations, see, e.g., \cite{GNR}, \cite{NY}, \cite{JS}, \cite{Sak}. The equations \tht{0.7} turn out to be a special case of the dPV equation of \cite{Sak}.
In this paper the discrete analogs of Painlev\'e equation are derived from a purely algebraic geometric construction. We refer the reader to \cite{Sak} for a further discussion of the subject. 

Discrete Painlev\'e I and II equations have appeared in the physics literature earlier, see e.g. \cite{BK}, \cite{FIK}, \cite{PS}. One of the main points of the present paper is that dPV also arises in a concrete mathematical/physical model.

Let us also point out that the results of \cite{BD} in the continuous situation suggest that $\q$ is a natural candidate for the $\tau$-function of \tht{0.7}. However, no general definition of a $\tau$-function in the geometric setting of \cite{Sak} is available at this moment. 

The proofs of both theorems are based on the Riemann--Hilbert approach to discrete integrable operators developed in \cite{Bor2}. Both $p_k^{(\eta)}$ and $q_k^{(z,z',\xi)}$ are represented as $\det(1-K|_{\{k,k+1,\dots\}})$, where $K$ is either the {\it discrete Bessel kernel} of \cite{BOO}, \cite{J2}, or the hypergeometric (discrete ${}_2F_1$) kernel
of \cite{BO2}. 

Using the results of \cite{Bor2}, we reduce the computation of the Fredholm determinant to solving a certain discrete Riemann--Hilbert problem (DRHP, for short). The jump matrices of the DRHPs thus obtained have particularly simple form. This allows us to derive a Lax pair of difference equations for the solutions $m_k(\ze)$ of these DRHPs which has the form
$$
m_{k+1}(\ze)=A(\ze)m_k(\ze),\qquad m_k(\ze-1)=B(\ze)m_k(\ze)C(\ze)
$$
with some unknown rational matrices $A(\ze)$ and $B(\ze)$ and a known rational diagonal matrix $C(\ze)$. The consistency relations for these two equations lead to discrete Painlev\'e equations on the matrix elements of $A$ and $B$. Additional arguments are needed to express the Fredholm determinants through these matrix elements.

The continuous variant of the same scheme has been worked out in \cite{BD}. There the corresponding matrix $m_t(\ze)$ solves a RHP with a jump matrix which can be conjugated to a piecewise constant one. This leads to the equations
$$
\frac{\partial m_t(\ze)}{\partial t}=\Cal A(\ze)m_t(\ze),\qquad
\frac{\partial m_t(\ze)}{\partial \ze}=\Cal B(\ze)m_t(\ze)+m_t(\ze)\Cal C(\ze)
$$
with known $\Cal C$ and unknown rational $\Cal A$ and $\Cal B$.
They form a Lax pair for an appropriate Painlev\'e equation, and the Fredholm determinant in question is the $\tau$-function of the isomonodromy deformation associated with the RHP. Similar ideas in the continuous setting were also used in \cite{Pal}, 
\cite{HI}, \cite{DIZ}, \cite{KH}. 

We expect that our approach can also be applied to a variety of other discrete integrable kernels, in particular, to the Christoffel--Darboux 
kernels for discrete orthogonal polynomials of the Askey--Wilson scheme. The case of Charlier polynomials, also related to the longest increasing subsequences in random words, will be worked out in the subsequent paper \cite{BB}. It leads to the dPIV equation of \cite{Sak}.

I am very grateful to Percy Deift, David Kazhdan, and Grigori Olshanski for interesting and helpful discussions. I would also like to thank
Craig Tracy for his letter \cite{T} which initiated my interest in the subject,
Jinho Baik and Pierre van Moerbeke for keeping me informed about their work, Masatoshi Noumi for referring me to \cite{Sak}, and Hidetaka Sakai for useful comments about his paper.

This research was partially conducted during the period the author served as a Clay Mathematics Institute Long-Term Prize Fellow. This work was also partially supported by the NSF grant DMS-9729992.

\head 1. Integrable operators and discrete Riemann--Hilbert problems
\endhead

In this section we give a summary of results proved in \cite{Bor2, \S4}.

Let $\y$ be a discrete locally finite subset of $\C$. We call an operator $K$ acting in $\ell^2(\y)$ {\it integrable} if its matrix has the form
$$
K(x,y)=\cases \dfrac {\sum_{j=1}^N F_j(x)G_j(y)}{x-y},&\quad x\ne y,\\
k(x),&\quad x=y,
\endcases
\tag 1.1
$$
for some complex--valued functions $F_j,\ G_j$, $j=1,\dots,N$, and $k$ on $\y$. 

Integrable operators as a distinguished class were first singled out in 
the continuous setting in \cite{IIKS}, see also \cite{D1}. The definition
in the discrete setting was given in \cite{Bor2}.

We will assume that

$\bullet$ $F_j,G_j\in\ell^2(\y)$ for all $j=1,\dots,N$, and
$$
\sum_{j=1}^N F_j(x)G_j(x)=0,\quad x\in\y.
\tag 1.2
$$

$\bullet$ $k$ is a bounded function on $\y$ which is also bounded away from $1$. That is, 
$$
\inf_{x\in\y}|k(x)-1|>0.
$$

$\bullet$
The operator
$$
(Th)(x)=\sum_{x'\in\y,\, x'\ne x}\frac{h(x')}{x-x'}
$$
is a bounded operator in $\ell^2(\y)$. This always holds if, for example, $\y$ is a subset of a one--dimensional lattice in $\C$. 

It is not hard to show that under these assumptions, $K$ is a bounded operator in $\ell^2(\y)$.

Our goal is to explain how the operator $(1-K)^{-1}$ (if it exists) can be expressed through a solution of a complex analytic problem which we call the {\it discrete Riemann--Hilbert problem} (DRHP, for short).

Let $w$ be a map from $\y$ to $\Ma(n,\C)$, where $n$ is a fixed integer.

We say that a matrix--valued function $m:\C\setminus\y\to \Ma(n,\C)$ 
with simple poles at the points $x\in\y$ is a solution of the DRHP $(\y,w)$ if the following conditions are satisfied 
$$
\align
&\bullet\quad  m(\zeta)\text{ is analytic in }\Bbb C\setminus \y,\\
&\bullet\quad 
\r m(\ze)=\lim_{\zeta\to x}\left(m(\zeta)w(x)\right),\quad x\in\y.
\endalign
$$
By analogy with continuous Riemann--Hilbert problems, we call $w(x)$ the {\it jump matrix}.

We say that $m$ satisfies the {\it normalized} DRHP $(\y,w)$ if, in addition to the conditions above, $ m(\zeta)\to I$ as  $\zeta\to \infty$. Here $I$ is the $n\times n$ identity matrix.

If the set $\y$ is infinite, the last condition must be made more precise. Indeed, a function with poles accumulating at infinity cannot have a limit at infinity. One way to make the condition precise is to require uniform asymptotics on a sequence of expanding contours, for example, on a sequence of circles $|\ze|=a_k$, $a_k\to+\infty$. 

Whenever below we give an asymptotic behavior of a function with poles in $\y$ at infinity, we mean that there exists a sequence of expanding contours such that the distance from these contours to the set $\y$ is bounded away from zero, and	 the function has uniform asymptotics on these contours.  

Let us introduce column--vectors
$$
F=(F_1,\dots,F_N)^t,\quad G=(G_1,\dots,G_N)^t.
$$
Then (1.2) can be rewritten as $F^t(x)G(x)=G^t(x)F(x)=0$.

\proclaim{Theorem 1.1}
Let $K$ be an integrable operator as defined above, and assume that the operator $(1-K)$ is invertible. Then

$(i)$ There exists a unique solution $m_\y$ of the normalized DRHP $(\y,w)$ where
$$
w(x)=(1-k(x))^{-1}F(x)G^t(x)\in\Ma(N,\C).
$$
Moreover, $\det m_\y\equiv 1$.

$(ii)$ The matrix of the operator $R=(1-K)^{-1}-1=K(1-K)^{-1}$  has the form
$$
R(x,y)=\cases \dfrac {\sum_{j=1}^N \F_j(x)\G_j(y)}{x-y}\,,
&\quad x\ne y,
\\
\dfrac {k(x)+r(x)}{1-k(x)}\,,
&\quad x=y,
\endcases
\tag 1.3
$$
where, with the notation $\F=(\F_1,\dots,\F_N)^t$, $\G=(\G_1,\dots,\G_N)^t$,
$$
\gathered
\F(x)=(1-k(x))^{-1}\cdot \lim\limits_{\ze\to x}m_\y(\ze)F(x),\\ \G(x)=(1-k(x))^{-1}\cdot \lim\limits_{\ze\to x}m_\y^{-t}(\ze)G(x),
\\
r(x)=\G^t(x)\lim_{\ze\to x}\left(m'_\y(\ze)F(x)\right).
\endgathered
\tag 1.4
$$
Here $m'_\y(\ze)=dm_\y(\ze)/d\ze$.
\endproclaim
\demo{Proof} Follows from Proposition 4.3 and Remark 4.2 in \cite{Bor2}. Note the difference in notation: we used $\x$, $L$, $f$, $g$, and $K=L(1+L)^{-1}$ in \cite{Bor2} instead of $\y$, $K$, $F$, $G$ and $R=K(1-K)^{-1}$ here. The reason for switching the notation will become clear in the next section.\qed 
\enddemo

\head 2. Simpler DRHP for a class of integrable operators
\endhead
It turns out that for many integrable operators of interest, the DRHP from Theorem 1.1(i) is rather complicated. In this section we show how to reduce the computation of the inverse operator $(1-K)^{-1}$ and $\det(1-K)$ to a much simpler DRHP for a certain subclass of the class of integrable operators.

Our main assumption is that there exists a locally finite set $\x$, $\x\supset\y$, vector--valued functions
$$
f=(f_1,\dots,f_N)^t,\quad g=(g_1,\dots,g_N)^t
$$ 
on $\x$, and a matrix--valued 
function 
$m_\x:\C \setminus\x\to \C$
such that 

(1) $f^t(x)g(x)\equiv 0$ on $\x$;

(2) $m_\x$ solves the DRHP $(\x,-fg^t)$, not necessarily normalized;

(3) $\det m_\x\equiv 1$;

(4) for any $x\in\y$
$$
\gathered
F(x)=\lim_{\ze\to x} m_\x(\ze) f(x),\quad 
G(x)=\lim_{\ze\to x} m_\x^{-t}(\ze) g(x)\\
k(x)=G^t(x)\lim_{\ze\to x} \left(m'_\x(\ze) f(x)\right).
\endgathered
$$

Note that we do not impose any boundedness/decay conditions on $f_j$ and $g_j$. 

There are at least two general situations when this assumption is satisfied.

\example{Situation 2.1} There exists an integrable operator $L$ on $\x$ with the matrix
$$
L(x,y)=\cases \dfrac {\sum_{j=1}^N f_j(x)g_j(y)}{x-y},&\quad x\ne y,\\
0,&\quad x=y.
\endcases
$$
Then if we set $K=L(1+L)^{-1}$, assuming that $-1$ is not in the spectrum of $L$, we will have the main assumption above satisfied by Theorem 1.1 (or Proposition 4.3 in \cite{Bor2}). Note that $m_\x$ will then satisfy the normalized DRHP. 
\endexample

\example{Situation 2.2} The matrix $K$ has the form
$$
K(x,y)=\cases \alpha(x)\beta(y)\,
\dfrac {\phi(x)\psi(y)-\psi(x)\phi(y)}{x-y}\,,&\quad x\ne y,
\\
\alpha(x)\beta(x)\left(\phi'(x)\psi(x)-\psi'(x)\phi(x)\right),
&\quad x=y,
\endcases
$$
where $\alpha,\beta$ are some functions defined on $\y$, and $\phi,\psi$ are {\it entire} functions. Note that the expression for $K(x,x)$ is obtained from that for $K(x,y)$ by formal limit transition $y\to x$.

Assume that there exist meromorphic functions $\hat\phi$ and $\hat\psi$ with simple poles, such that for any $x\in\y$ we have
$$
\r\hat\phi(\ze)=\alpha(x)\beta(x)\phi(x), \quad
\r\hat\psi(\ze)=\alpha(x)\beta(x)\psi(x),
$$
and $\phi\hat\psi-\psi\hat\phi\equiv 1$ on $\C$. 

Let us denote by $\x$ the union of $\y$ and the set of poles of the functions $\hat\phi$ and $\hat\psi$. Assume that we can extend the functions $\alpha$ and $\beta$ to $\x$ in such a way that the residue conditions above hold for $x\in\x$.
 
Then we can satisfy the main assumption (1)--(4) by setting 
$$
\gathered
f=(\alpha,0)^t,\quad g=(0,-\beta)^t,\qquad F=(\alpha\phi,\alpha\psi)^t,\quad
G=(\beta\psi,-\beta\phi)^t,\\
m_\x=\bmatrix \phi&\hat\phi\\ \psi&\hat\psi\endbmatrix,\quad
m_\x^{-t}=\bmatrix \hat\psi&-\psi\\ \hat\phi&\phi\endbmatrix.
\endgathered
$$
\endexample

Situation 2.1 comes up when $K$ is the correlation kernel for a determinantal point process defined as an {\it L--ensemble}, see \cite{BO4, \S5}, and also \cite{BO2}, \cite{BOO}, and \cite{Bor2} for concrete examples.

Situation 2.2 suits (restrictions of) Christoffel--Darboux kernels for classical discrete orthogonal polynomials and some limits of such kernels. The functions $\hat\phi$ and $\hat\psi$ are then the so-called ``functions of the second kind'', $\x$ is the orthogonality set, and the product $\alpha\beta$ is equal to the weight function.

Denote $\z=\x\setminus\y$.

\proclaim{Theorem 2.3} Under the main assumption $(1)-(4)$ above

$(i)$ There exists a unique solution $m_{\z}$ of the DRHP $(\z,-fg^t)$ satisfying the condition $m_{\z}m^{-1}_{\x}\to I$ as $\ze\to\infty$. Moreover, $\det m_\z\equiv 1$.

$(ii)$ The matrix of the operator $R=(1-K)^{-1}-1=K(1-K)^{-1}$  has the form
$$
R(x,y)=\cases \dfrac {\sum_{j=1}^N \F_j(x)\G_j(y)}{x-y},&\quad x\ne y,
\\
g^t(x)m_\z^{-1}(x)m_\z'(x)f(x),&\quad x=y,
\endcases
\tag 2.1
$$
where
$$
\F(x)=m_\z(x)f(x),\quad \G(x)=m_\z^{-t}(x)g(x).
\tag 2.2
$$
\endproclaim
\demo{Comments} 1. Since $\det m_\x\equiv\det m_\z\equiv 1$, the inverse matrices $m_\x^{-1}$ and $m_\z^{-1}$ are well--defined outside $\x$ and $\z$ respectively.

2. The formulas (2.1) and (2.2) hold for $x,y\in\y$, and $m_\z$ is analytic around such points because $\y\cap\z=\varnothing$.

3. The part of Theorem 2.3 which will be employed later on, is the formula for the diagonal values $R(x,x)$. The reason is the relation (assume that $K$ is a trace class operator)
$$
1+R(x,x)=\frac{\det(1-K_x)}{\det(1-K)},
$$ 
where $K_x$ is the restriction of the matrix $K$ to $\y\setminus x$. 

4. The DRHP of Theorem 2.3 in a number of interesting examples turns out to be much simpler than that of Theorem 1.1. The reason is that the functions $\{f_j,g_j\}$ are often elementary (or even constant), while the functions $\{F_j,G_j\}$ are commonly expressed through classical special functions.
\enddemo

Before proceeding to the proof of Theorem 2.3, let us prove the following
lemma.

\proclaim{Lemma 2.4} Let $X,Y$, and $Z$ be locally finite subsets of $\C$ such that $X=Y\sqcup Z$. Let $w_X$ be an arbitrary matrix--valued function on $X$ such that for any $x\in X$, $w_X(x)$ is nilpotent of rank 1. Let $m_X$ be a solution of the DRHP $(X,w_X)$ such that $\det m_X\equiv 1$. 

For any $x\in X$, denote by $B(x)$ the constant term in the Laurent expansion of $m_X$ near $x$. Then $\det B(x)\ne 0$. 
Furthermore, if $m_Z$ is a solution of the DRHP $(Z,w_X)$, then $m_Y\equiv m_Z m_X^{-1}$ solves the DRHP $(Y,w_Y)$, where
$$
w_Y(x)=-B(x)w_X(x)B^{-1}(x).
\tag 2.3
$$

Conversely, if $m_Y$ is a solution of the DRHP $(Y,w_Y)$, then $m_Z\equiv m_Ym_X$ solves the DRHP $(Z,w_X)$.
\endproclaim
\demo{Proof} Take any $x\in\x$. The residue condition implies that in the neighborhood of $x$ we have
$$
m_X(\ze)=B(x)\left(I+\frac{w_X(x)}{\ze-x}+(\ze-x)C(\ze)\right)
\tag 2.4
$$
where $C(\ze)$ is analytic in the neighborhood of $x$, cf. \cite{Bor2, Lemma 4.4}. Then $\det m_X\equiv 1$ implies $\det B(x)\ne 0$.

Since $w_X$ is of rank 1, there exist column--vectors $u$ and $v$ such 
that $w_X=uv^t$. The condition $w_X^2=0$ implies $u^tv=v^tu=0$. Then
$$
\det\left(I+\frac{w_X(x)}{\ze-x}+(\ze-x)C(\ze)\right)=
1-v^t(x)C(x)u(x)+O(\ze-x).
$$
Hence, $1-v^t(x)C(x)u(x)=(\det B(x))^{-1}\ne 0$.

Denote $A(\ze)=I+(\ze-x)C(\ze)$. Clearly, $A(\ze)$ is invertible when $|\ze-x|$ is small enough, and $A^{-1}(\ze)=I-(\ze-x)C(\ze)+O\left((\ze-x)^2\right)$. Hence
$$
\ze-x+v^t(x)A^{-1}(\ze)u(x)=(\ze-x)\left(1-v^t(x)C(x)u(x)\right)
+O\left((\ze-x)^2\right)
$$
is not equal to 0 if $\ze\ne x$ and $|\ze-x|$ is small.  

In the neighborhood of $x$ we have 
$$
\gather
\left(I+\frac{w_X(x)}{\ze-x}+(\ze-x)C(\ze)\right)^{-1}
=\left(A(\ze)+\frac{u(x)v^t(x)}{\ze-x}\right)^{-1}
\\
=
\left(I-\frac{A^{-1}(\ze)u(x)v^t(x)}{\ze-x+v^t(x)A^{-1}(\ze)u(x)}\right)
A^{-1}(\ze) 
\tag 2.5
\\
=
\left(I-\frac{u(x)v^t(x)}{(\ze-x)(1-v^t(x)C(x)v(x))}+O(1)\right)
A^{-1}(\ze)\\
=\left(I-\frac{\det B(x)u(x)v^t(x)}{\ze-x}+O(1)\right)A^{-1}(\ze)=-\frac{\det B(x)u(x)v^t(x)}{\ze-x}+O(1).
\endgather
$$

Thus, using \tht{2.4} wee that for any $x\in\x$
$$
\r m_X^{-1}(\ze)=-\det B(x)\cdot u(x)v^t(x)B^{-1}(x).
$$

On the other hand, relations \tht{2.4}, \tht{2.5} imply
$$
\gathered
m_X^{-1}(\ze) B(x)u(x)v^t(x)B^{-1}(x)
=\frac{(\ze-x)A^{-1}(\ze)}
{\ze-x+v^t(x)A^{-1}(\ze)u(x)}\,u(x)v^t(x)B^{-1}(x)\\
=\det B(x)\cdot u(x)v^t(x)B^{-1}(x)+O(\ze-x).
\endgathered
$$

Hence, for $x\in X$,
$$
\r m_X^{-1}(\ze)=-\lim_{\ze\to x} 
\left(m_X^{-1}(\ze)\,(B(x)w_X(x)B^{-1}(x))\right).
$$

Now, if $x\in Y$ then $m_Z$ is analytic near $x$, and for 
$m_Y=m_Zm_X^{-1}$ we have
$$
\r m_Y(\ze)=-\lim_{\ze\to x} \left(m_Y(\ze)\,(B(x)w_X(x)B^{-1}(x))\right)
$$
as required. For $x\in Z$, both $m_X$ and $m_Z$ satisfy the same residue condition at $x$, which implies that $m_Y$ is analytic near $x$, cf. \cite{Bor2, Lemma 4.5}. This concludes the proof of the first statement of the lemma.

To prove the converse statement it suffices to verify that $m_Y m_X$ is analytic near any $x\in Y$. But we have just proved that $m^{-1}_X$ satisfies the same residue condition as $m_Y$ does. Thus, their ratio is analytic.\qed
\enddemo

\demo{Proof of Theorem 2.3} Let $m_\y$ be as in Theorem 1.1(i). Set 
$m_\z=m_\y m_\x$. Clearly, $\det m_\z\equiv 1$. We are going to show that for $x\in\y$
$$
(1-k(x))^{-1}F(x)G^t(x)=B_\x(x)f(x)g^t(x)B_\x(x)^{-1},
\tag 2.6
$$
where $B_\x(x)$ is the constant term in the Laurent expansion of $m_\x(\ze)$ 
near $x$. Then Lemma 2.4 will imply that $m_\z$ satisfies the DRHP
$(\z,-fg^t)$.\footnote{The match with Lemma 2.4 is established by
$X=\x$, $Y=\y$, $Z=\z$, $w_X=-fg^t$, $w_Y=(1-k)^{-1}FG^t$, $u=f$, $v=-g$.}

Condition (2) of the main assumption implies that near $x\in\x$, $m_\x(\ze)$ has the form
$$
m_\x(\ze)=B_\x(x)\left(I-\frac{f(x)g^t(x)}{\ze-x}+(\ze-x)C_\x(\ze)\right),
$$
where $C_\x(\ze)$ is analytic in the neighborhood of $x$, cf. \tht{2.4}. Then condition (1) of the main assumption implies
$$
F(x)=\lim_{\ze\to x}m_\x(\ze)f(x)=B_\x(x)f(x).
\tag 2.7
$$

Similarly to the proof of Lemma 2.4, 
setting $A_\x(\ze)=I+(\ze-x)C_\x(\ze)$, we have
$$
\gathered
m_\x^{-t}(\ze)g(x)=B_\x^{-t}(x)A_\x^{-t}(\ze)\left(g(x)+
\frac{g(x)f^t(x)A_\x^{-t}(\ze)g(x)}
{\ze-x-g^t(x)A_\x^{-1}(\ze)f(x)}\right)\\
=\frac{(\ze-x)B_\x^{-t}(x)A_\x^{-t}(\ze)g(x)}
{{\ze-x-g^t(x)A_\x^{-1}(\ze)f(x)}}=\det B_\x(x)\cdot B_\x^{-t}(x)g(x)+O(1),
\endgathered
$$
as $\ze\to x$. That is, 
$$
G(x)=\lim_{\ze\to x}m_\x^{-t}(\ze)g(x)=\det B_\x(x)\cdot B_\x^{-t}(x)g(x).
\tag 2.8
$$ 

Finally, 
$$
\gathered
k(x)=G^t(x)\lim_{\ze\to x}\left(m'_\x(\ze)f(x)\right)=G^t(x)B_\x(x)C_\x(x)f(x)\\
=\det B_\x(x)g^t(x)C_\x(x)f(x)=\det B_\x(x)
\left((\det B_\x(x))^{-1}-1\right)=1-\det B_\x(x).
\endgathered
\tag 2.9
$$
Then \tht{2.6} follows from \tht{2.7}--\tht{2.9}.

The uniqueness of $m_\z$ follows from the following general argument. If 
$n_\z$ is another solution of $(\z,-fg^t)$ with the same asymptotics at
infinity then $n_\z m_\z^{-1}$ has no singularities and tends to $I$ at infinity. By Liouville's theorem, $n_\z\equiv m_\z$. This concludes the 
proof of (i).

Let us proceed to (ii). Thanks to \tht{2.6}, locally near any $x\in Y$ we can write $m_\y$ in the form
$$
m_\y(\ze)=B_\y(x)B_\x(x)\left(I+\frac{f(x)g^t(x)}{\ze-x}+(\ze-x)C_\y(\ze)
\right)B_\x^{-1}(x),
$$
where $B_\y(x)$ is a nondegenerate matrix, and $C_\y(\ze)$ is analytic near $x$. Then, similarly to \tht{2.7}, \tht{2.8}, using \tht{2.7}--\tht{2.9}, we get
$$
\gathered
\F(x)=(1-k(x))^{-1}\cdot\lim_{\ze\to x}m_\y(\ze)F(x)=(\det B_\x(x))^{-1}\cdot B_\y(x) B_\x(x) f(x),\\
\G(x)=(1-k(x))^{-1}\cdot\lim_{\ze\to x}m_\y^{-t}G(x)=\det\left(B_\y(x)\right)\cdot
(B_\y B_\x)^{-t}(x)g(x).
\endgathered
\tag 2.10
$$

Further, near $x\in Y$ we have
$$
\gathered
m_\z(\ze)=m_\y(\ze)m_\x(\ze)
=B_\y(x)B_\x(x)\\ \times
\left(I+f(x)g^t(x)C_\x(\ze)-C_\y(\ze)f(x)g^t(x)+
(C_\x(\ze)+C_\y(\ze))(\ze-x)
\right).
\endgathered
\tag 2.11
$$
Thus,
$$
\gathered
m_\z(x)f(x)=(1+g^t(x)C_\x(x)f(x))\cdot B_\y(x)B_\x(x)f(x)=\F(x).
\endgathered
\tag 2.12
$$

It is easy to verify the following equality (remember that
$1+g^tC_\x f=(\det B_\x)^{-1}$ and 
$1-g^tC_\y f=(\det B_\y)^{-1}$)
$$
\gathered
m_\z^{-1}B_\y B_\x=
\left(I+fg^tC_\x-C_\y fg^t)\right)^{-1}\\
=I-\det B_\x\cdot fg^tC_\x+\det B_\y\cdot 
C_\y fg^t-\det(B_\x B_\y)\cdot fg^tC_\x C_\y fg^t, 
\endgathered
$$
where we have omitted the argument $x\in Y$ of all the functions above.
Hence,
$$
\gathered
m_\z^{-t}(x)g(x)=(B_\y B_\x)^{-t}(x)(1+\det B_\y(x)\,(g^tC_\y f)(x))\cdot g(x)\\ =\det B_\y(x)\cdot (B_\y B_\x)^{-t}(x)g(x)=\G(x).
\endgathered
\tag 2.13
$$

The formula \tht{2.1} for the nondiagonal entries of $R$ follows from \tht{1.3}, \tht{2.12} and \tht{2.13}.

To prove \tht{2.1} for diagonal entries we need to evaluate $r(x)$ introduced in \tht{1.4}. We have
$$
\lim_{\ze\to x} (m'_\y(\ze)F(x))=B_\y(x) B_\x(x) C_\y(x) f(x).
$$
Thus, with the help of \tht{2.10},
$$
r(x)=\G^t(x)\lim_{\ze\to x}\left(m'_\y(\ze)F(x)\right)=
\det B_\y(x)\cdot (g^tC_\y f)(x)=
\det B_\y(x) -1.
$$
Then 
$$
R(x,x)=\frac{k(x)+r(x)}{1-k(x)}=\frac{\det B_\y(x)}{\det B_\x(x)}-1.
$$

On the other hand, \tht{2.11} and \tht{2.13} imply
$$
\gathered
g^t m_\z^{-1} m'_\z f=\det B_\y\cdot g^t (C_\x+C_\y)f\\=\det B_\y\cdot
((\det B_\x^{-1}-1)+(1-\det B_\y^{-1}))=\frac{\det B_\y(x)}{\det B_\x(x)}-1.\qed
\endgathered
$$

\enddemo

\head 3. Discrete Bessel kernel and dPII 
\endhead

In this section we will apply the general formalism of \S2 to derive the discrete Painlev\'e II equation (dPII, for short) for the Fredholm determinant of the {\it discrete Bessel kernel}.

The discrete Bessel kernel plays an important role in the asymptotic
analysis of the Plancherel measures on the symmetric groups. It was derived independently in \cite{BOO} and \cite{J2}, see also \cite{Bor2}.
We define it as follows.

Let $\Z'=\Z+\frac 12=\{\pm\frac 12,\pm\frac 32,\dots\}$. For $x,y\in \Z'$
set
$$
K(x,y)=\eta\,\frac{J_{x-\frac 12}(2\eta)J_{y+\frac 12}(2\eta)-
J_{y-\frac 12}(2\eta)J_{x+\frac 12}(2\eta)}{x-y},
\tag 3.1'
$$
where $\eta>0$ is a parameter, $J_\nu(\,\cdot\,)$ is the $J$--Bessel function, and $K(x,x)$ is defined by the L'Hospital rule:
$$
K(x,x)=\eta\left(
\frac{\partial J_{x-\frac 12}(2\eta)}{\partial x}\,J_{x+\frac 12}(2\eta)-
J_{x-\frac 12}(2\eta)\,\frac{\partial J_{x+\frac 12}(2\eta)}{\partial x}\right).
\tag 3.1''
$$

Note that only the Bessel functions with {\it integral} indices enter the formula. 

For any $s\in \Z'$, denote by $K_s$ the operator in $\ell^2(\{s,s+1,\dots\})$ defined by the restriction of $K$ to
$\{s,s+1,\dots\}\times \{s,s+1,\dots\}$. It can be shown that $K_s$ is a positive trace class operator.\footnote{This immediately follows from
the fact that $K_s$ is a symmetric correlation kernel for a determinantal point process which has finitely many particles almost surely, see \cite{So, Theorem 4} for a general theorem and \cite{BOO} for the description of the point process.}

We will be interested in the Fredholm determinants
$$
D_s=\det(1-K_s).
$$
From the probabilistic interpretation of the kernel $K$ it immediately follows that $D_s=0$ if $s<0$.\footnote{Vanishing of $D_s$ is equivalent to the statement that for any partition $\lambda=(\lambda_1,\lambda_2,\dots)$, the set 
$\{\lambda_i-i+\frac 12\}\cap \{s,s+1,\dots\}$ is nonempty, see \cite{BOO}.} Thus, from now on we assume that $s>0$. Then $D_s>0$. 

It is worth noting that $D_s$ can also be interpreted as a Toeplitz determinant, see \cite{BOk, \S4}. Namely,
$$
D_s=\exp(-\eta^2)\cdot \det[I_{i-j}(2\eta)]_{i,j=1,\dots,s-\frac 12}\,
. 
\tag 3.2
$$
Here $I_\nu(\,\cdot \,)$ is the $I$--Bessel function. This formula also follows from \cite{G}, \cite{BOO}, \cite{J2}. The symbol of the Toeplitz determinants above is equal to
$$
\sum_{k\in\Z}I_k(2\eta)\ze^k=\exp(\eta(\ze+\ze^{-1})).
$$

The identity \tht{3.2} can be generalized to Toeplitz determinants with an arbitrary symbol \cite{BOk}, and even to block Toeplitz determinants 
\cite{BW}, see \cite{Bot} for a simple proof. 

Note that in the Introduction we used the notation $p^{(\eta)}_{s-\frac 12}$ for $D_s$.

The discrete Bessel kernel fits into both situations described in the beginning of \S2. We will use Situation 2.1, but let us first show how 
Situation 2.2 works in this case. Naturally, set $\y=\{s,s+1,\dots\}$.

\example{Example 3.1 (a realization of Situation 2.2)}
We set $\x=\Z'$, and
$$
\gathered
\alpha\equiv 1,\quad \beta\equiv 1,\quad 
\phi(\ze)=J_{\ze-\frac 12}(2\eta),
\quad \psi(\ze)=\eta J_{\ze+\frac 12}(2\eta),\\
\hat\phi(\ze)=-\frac{\pi}{\cos\pi\ze}\,{J_{-\ze+\frac 12}(2\eta)},\quad 
\hat\psi(\ze)=\frac{\pi \eta}{\cos\pi \ze}\,{J_{-\ze-\frac 12}(2\eta)}.
\endgathered
$$

Then the residue conditions follow from the relation $J_{-n}=(-1)^n J_n$, $n\in\Z$. The equality $\phi\hat\psi-\psi\hat\phi\equiv 1$ is the well--known identity
$$
J_{\ze-\frac 12}(2\eta)J_{-\ze-\frac 12}(2\eta)+
J_{\ze+\frac 12}(2\eta)J_{-\ze+\frac 12}(2\eta)
=\frac{\cos\pi\ze}{\pi\eta}\,.
$$

Note also that the matrix 
$$
m_\x(\ze)=\bmatrix J_{\ze-\frac 12}(2\eta)& -\frac{\pi}{\cos\pi\ze}\,{J_{-\ze+\frac 12}(2\eta)}\\
\eta J_{\ze+\frac 12}(2\eta)&
\frac{\pi \eta}{\cos\pi \ze}\,{J_{-\ze-\frac 12}(2\eta)}\endbmatrix
$$ 
satisfies the equation
$$
m_\x(\ze-1)=\frac 1\eta\bmatrix \ze-\frac 12 &-1\\
\eta&0\endbmatrix m_\x(\ze).
$$

The only disadvantage of this realization is that $m_\x(\ze)$ does
not tend to $I$ as $\ze\to\infty$. However, one can very well use it to 
analyze the inverse operators $(1-K_s)^{-1}$  and the Fredholm 
determinants $D_s$.
\endexample

All the details of how Situation 2.1 is applied to the discrete Bessel kernel are explained in \cite{Bor2}. Here we just state the results
that we need. 

We set $\x=\Z'$. The matrix $m_\x$ has the form
$$
m_\x(\ze)=\sqrt{\eta}\bmatrix J_{\ze-\frac 12}(2\eta)&J_{-\ze+\frac 12}(2\eta)\\
-J_{\ze+\frac 12}(2\eta)&J_{-\ze-\frac 12}(2\eta)\endbmatrix\bmatrix {\eta^{-\ze}\Gamma(\ze+\frac 12)}&0\\ 0 &{\eta^{\ze}\Gamma(-\ze+\frac 12)}\endbmatrix.
$$

It solves the normalized DRHP $(\x,-fg^t)$, where $f=(f_1,f_2)^t$, $g=(g_1,g_2)^t$,
$$
\gathered
f_1(x)=\cases \frac{\eta^x}{\Gamma(x+\frac 12)},&x\in\Z'_+,\\
0,&x\in\ \Z'_-,\endcases\quad
f_2(x)=\cases 0,&x\in\Z'_+,\\
\frac{\eta^{-x}}{\Gamma(-x+\frac 12)},&x\in\Z'_-,\endcases
\\
g_1(x)=\cases 0,&x\in\Z'_+,\\
\frac{\eta^{-x}}{\Gamma(-x+\frac 12)},&x\in\Z'_-, 
\endcases\quad
g_2(x)=\cases \frac{\eta^x}{\Gamma(x+\frac12)},&x\in\Z'_+,
\\0,&x\in\Z'_-.\endcases
\endgathered
$$
The jump matrix $w_\x(x)=-f(x)g(x)^t$ has the form
$$
w_\x(x)=\cases \bmatrix 0&-\frac{\eta^{2x}}{\Gamma^2(x+\frac 12)}\\0&0\endbmatrix,&x\in\Z'_+,\\
\bmatrix 0&0\\-\frac{\eta^{-2x}}{\Gamma^2(-x+\frac 12)}&0\endbmatrix,&x\in\Z'_-.
\endcases
\tag 3.3
$$

We proceed to examine the unique solution $m_\z$ of the normalized DRHP
$(\z,w_\x)$, where 
$$
\z=\x\setminus\y=\Z\setminus\{s,s+1,\dots\}=\{\dots,s-2,s-1\},
$$
see Theorem 2.3(i). It is more convenient to redenote $m_\z$ by $m_s$, because we
will be working with solutions corresponding to different values of $s$. 
We also denote $\z_s=\{\dots,s-2,s-1\}$.
Recall that $\det m_s\equiv 1$. 

We aim at proving the following
\proclaim{Proposition 3.2 (Lax pair)} For any $s\in\Z'_+$ there exist
a constant nilpotent matrix $A_s$,
$$
A_s=\bmatrix p_s&q_s\\r_s&-p_s\endbmatrix,\quad p_s^2=-r_sq_s,
$$
and constant $a_s,b_s$, $a_sb_s=1$, such that
$$
\gather
m_{s+1}(\ze)=\left(I+\frac {A_s}{\ze-s}\right)m_s(\ze), \tag 3.4\\
m_s(\ze-1)=\bmatrix \eta^{-1}(\ze-\frac 12-p_s)&a_s\\-b_s&0
\endbmatrix m_{s+1}(\ze)\bmatrix \eta(\ze-\frac 12)^{-1}&0\\0& \eta^{-1}(\ze-\frac 12) \endbmatrix.
\tag 3.5
\endgather
$$
\endproclaim
\demo{Proof} The first equation is almost obvious. Indeed, since $w_\x$
does not depend on $s$, we see that $m_s(\ze)$ and $m_{s+1}(\ze)$ satisfy the same residue condition on $\z_s$. However, $m_{s+1}$ has an extra pole at the point $\{s\}=\z_{s+1}\setminus\z_s$. Hence, the ratio
$m_{s+1}m_s^{-1}$ has only one pole at the point $\ze=s$. Denoting the
residue at this pole by $A_s$, we conclude that the function
$$
m_{s+1}(\ze)m_s^{-1}(\ze)-\frac {A_s}{\ze-s}
$$
is entire. Evaluating the asymptotics at $\ze=\infty$ we see that,
by Liouville's theorem, this function is identically equal to $I$, which
proves the first equation. Furthermore, since 
$\det m_s\equiv\det m_{s+1}\equiv 1$, we see that $\det(I+{A_s}/{(\ze-s)})\equiv 1$. This implies that $A_s$ is nilpotent.

To prove the second equation, let us first verify that
$$
m_s(\ze-1)\bmatrix \eta^{-1}(\ze-\frac 12)&0\\0& \eta(\ze-\frac 12)^{-1}
\endbmatrix m_{s+1}^{-1}(\ze)
\tag 3.6
$$
is an entire function. It is easy to see that all the factors above are analytic outside $\z_{s+1}$. Let $x\in\z_{s+1}$. Then we have
$$
\gathered
m_s(\ze-1)= H_1(\ze)\left(I+\frac{w_\x(x-1)}{\ze-x}\right),\\
m_{s+1}^{-1}(\ze)=\left(I-\frac{w_\x(x)}{\ze-x}\right)H_2(\ze),
\endgathered
$$
where $H_1,H_2$ are analytic and invertible near $x$. Hence, we need to prove that
$$
\left(I+\frac{w_\x(x-1)}{\ze-x}\right)\bmatrix \eta^{-1}(\ze-\frac 12)&0\\0& \eta(\ze-\frac 12)^{-1}\endbmatrix
\left(I-\frac{w_\x(x)}{\ze-x}\right)
$$
is analytic. For $x\ne \frac 12$ this is obvious from \tht{3.3}. For $x=\frac 12$ we have
$$
\gathered
\bmatrix 1&0\\-{\eta}(\ze-\frac 12)^{-1}&1\endbmatrix\bmatrix \eta^{-1}(\ze-\frac 12)&0\\0& \eta(\ze-\frac 12)^{-1}\endbmatrix
\bmatrix 1&{\eta}(\ze-\frac 12)^{-1}\\0&1\endbmatrix\\
=\bmatrix \eta^{-1}(\ze-\frac 12)&0\\-1& \eta(\ze-\frac 12)^{-1}\endbmatrix
\bmatrix 1&{\eta}(\ze-\frac 12)^{-1}\\0&1\endbmatrix=\bmatrix \eta^{-1}(\ze-\frac 12)&1\\-1& 0\endbmatrix
\endgathered
$$
which is analytic. Thus, \tht{3.6} is entire.\footnote{This is no longer true
if $s\le-\frac 12$. Then \tht{3.6} has a pole at $\ze=\frac 12$.}

The next step is to compute the asymptotics of \tht{3.6} at infinity.
From the general formula \cite{Bor2, (4.9)} it follows that 
$$
\gather
m_s(\ze)=I+\bmatrix \alpha_s&\beta_s\\ \gamma_s&\delta_s\endbmatrix\,\ze^{-1}+O(\ze^{-2}),
\\
m_{s+1}(\ze)=I+\bmatrix \alpha_{s+1}&\beta_{s+1}\\ \gamma_{s+1}&\delta_{s+1}\endbmatrix\,\ze^{-1}+O(\ze^{-2}),
\endgather
$$
as $\ze\to\infty$, with some constants $\alpha_{s},\alpha_{s+1},\dots,\delta_{s},\delta_{s+1}$. Hence, the asymptotics of \tht{3.6} has the form
$$
\gathered
\left(I+\bmatrix \alpha_s&\beta_s\\ \gamma_s&\delta_s\endbmatrix\,\ze^{-1}\right)
\bmatrix \eta^{-1}(\ze-\frac 12)&0\\0& 0
\endbmatrix\left(I-\bmatrix \alpha_{s+1}&\beta_{s+1}\\ \gamma_{s+1}&\delta_{s+1}\endbmatrix\,\ze^{-1})\right)+O(\ze^{-1})\\
=\eta^{-1}\bmatrix \ze-\frac 12 +\alpha_s-\alpha_{s+1}& -\beta_{s+1}\\
\gamma_s&0\endbmatrix +O(\ze^{-1}).
\endgathered
$$
Denote $a_s=-\eta^{-1}\beta_{s+1}$, $b_s=-\eta^{-1}\gamma_s$, $c_s=
\alpha_{s+1}-\alpha_{s}$. Then Liouville's theorem implies that \tht{3.6} is equal to 
$$
\bmatrix \eta^{-1}(\ze-\frac 12-c_s)&a_s\\-b_s&0 \endbmatrix.
$$
It remains to show that $c_s=p_s$ and $a_sb_s=1$. The second equality follows from the fact that the determinant of \tht{3.6} is equal to 1. 
To prove that $c_s=p_s$, let us now substitute \tht{3.4} into what we have just proved. We obtain
$$
\gathered
m_s(\ze-1)\bmatrix \eta^{-1}(\ze-\frac 12)&0\\0& \eta(\ze-\frac 12)^{-1}
\endbmatrix\\
=\bmatrix \eta^{-1}(\ze-\frac 12-c_s)&a_s\\-b_s&0 \endbmatrix
\left(I+(\ze-s)^{-1} \bmatrix p_s&q_s\\r_s&-p_s\endbmatrix\right)m_s(\ze).
\endgathered
$$
Comparing the asymptotics of the (1,1) entry of both sides, we conclude that $c_s=p_s$.\qed
\enddemo

\proclaim{Corollary 3.3 (Compatibility conditions)} For any $s\in\Z'_+$, we have 
$$
\gather
a_s=a_{s+1}+\eta^{-1}q_{s+1},\quad b_s=b_{s+1}+\eta^{-1}r_s, \tag 3.7\\
a_sr_s=-b_{s+1}q_{s+1}.\tag 3.8
\endgather
$$
Denote $a_sr_s$ by $w_s$. Then
$$
\gather
(p_s+p_{s+1})(w_s-\eta)=\left(s+\tfrac 12\right)w_s,\tag 3.9\\
p_{s+1}^2=w_sw_{s+1}.\tag 3.10
\endgather
$$
\endproclaim
\demo{Proof} Shifting $\ze$ by 1 in \tht{3.5} and substituting the right--hand side of \tht{3.5} into the right--hand side of \tht{3.4} yields
$$
\multline
m_{s+1}(\ze)=\left(I+\frac{A_s}{\ze-s}\right)\bmatrix 
\eta^{-1}(\ze+\frac 12 -p_s)&a_s\\-b_s&0\endbmatrix
\\ \times m_{s+1}(\ze+1)\bmatrix \eta(\ze+\frac 12)^{-1}&0\\
0&\eta^{-1}(\ze+\frac 12)\endbmatrix.
\endmultline
$$

On the other hand, shifting $s$ and $\ze$ by 1 in \tht{3.4} and \tht{3.5}, and substituting the right--hand side of \tht{3.4} into the right--hand side of \tht{3.5} gives
$$
\multline
m_{s+1}(\ze)=\bmatrix 
\eta^{-1}(\ze+\frac 12 -p_{s+1})&a_{s+1}\\-b_{s+1}&0\endbmatrix
\left(I+\frac{A_{s+1}}{\ze-s}\right)
\\ \times m_{s+1}(\ze+1)\bmatrix \eta(\ze+\frac 12)^{-1}&0\\
0&\eta^{-1}(\ze+\frac 12)\endbmatrix.
\endmultline
$$

Comparing these two relation, we conclude that
$$
\left(I+\frac{A_s}{\ze-s}\right)\bmatrix 
\eta^{-1}(\ze+\frac 12 -p_s)&a_s\\-b_s&0\endbmatrix=
\bmatrix 
\eta^{-1}(\ze+\frac 12 -p_{s+1})&a_{s+1}\\-b_{s+1}&0\endbmatrix
\left(I+\frac{A_{s+1}}{\ze-s}\right).
\tag 3.11
$$

This is the {\it consistency} or {\it compatibility relation} for the Lax pair \tht{3.4}, \tht{3.5}. 

If we were in a continuous situation, the consistency relation 
would have been obtained by cross--differentiation of the Lax pair equations.

Computing the asymptotics of the (1,2) and (2,1) elements of \tht{3.11} at $\ze=\infty$ yields \tht{3.7}.
Let us compute the residue of both sides of \tht{3.11} at $\ze=s$. This gives
$$
\bmatrix p_s& q_s\\r_s&-p_s\endbmatrix
\bmatrix 
\eta^{-1}(\ze+\frac 12 -p_s)&a_s\\-b_s&0\endbmatrix=
\bmatrix 
\eta^{-1}(\ze+\frac 12 -p_{s+1})&a_{s+1}\\-b_{s+1}&0\endbmatrix
\bmatrix p_{s+1}& q_{s+1}\\r_{s+1}&-p_{s+1}\endbmatrix.
$$

The (2,2) element of this equality is \tht{3.8}. The (1,2) element gives
$$
a_sp_s=\eta^{-1}(\ze+\tfrac 12-p_{s+1})q_{s+1}-a_{s+1}p_{s+1}.
$$
Multiplying both sides by $b_{s+1}$, we obtain (recall that
$a_{s+1}b_{s+1}=1$)
$$
b_{s+1}a_sp_s=-\eta^{-1}(\ze+\tfrac 12-p_{s+1})w_s-p_{s+1}.
\tag 3.12  
$$

Multiplying the first relation of \tht{3.7} by $b_{s+1}$ we see that
$a_sb_{s+1}=1-\eta^{-1}w_s$. Substituting this into \tht{3.12} yields 
\tht{3.9}. 

As for \tht{3.10}, we have 
$$
p_{s+1}^2=-q_{s+1}r_{s+1}=(-b_{s+1}q_{s+1})(a_{s+1}r_{s+1})=w_sw_{s+1}.
\qed
$$
\enddemo

\proclaim{Proposition 3.4 (dPII)} Assume that $w_s\ne 0$ for $s=\frac 12,\frac 32,\dots, S$. Pick $v_\frac 12$ so that $v_\frac 12^2=\eta^{-1}w_\frac 12$, and define
$$
v_s=\frac{p_s}{\eta v_{s-1}},\quad s=\tfrac 32,\dots, S+1.
$$
Then for $s=\tfrac 32,\dots, S$ we have
$$
v_{s-1}+v_{s+1}=\frac{(s+\frac 12)v_s}{\eta(v_s^2-1)}.
\tag 3.13
$$
Moreover, $w_s=\eta v_s^2$ for all $s=\frac 12,\dots,S+1$.
\endproclaim

The relation \tht{3.13} is called the {\it discrete Painlev\'e II equation}, see e.g. \cite{GNR}.

\demo{Proof} The relation $w_s=\eta v_s^2$ follows from \tht{3.10} and
the definition of $v_s$. The equation \tht{3.13} follows from \tht{3.9}, after we substitute $\eta v_s^2$ for $w_s$, $\eta v_{s-1}v_s$ 
for $p_s$, and $\eta v_s v_{s+1}$ for $p_{s+1}$.\qed
\enddemo

\head 4. Fredholm determinant and dPII
\endhead

In this section we show how the Fredholm determinant $D_s$ is related
to the solution $v_s$ of the dPII introduced in Proposition 3.4.

We will work under the assumption that $w_s\ne 0$ for all $s\in\Z'_+$.
As we will see in the next section, this assumption holds for generic
values of $\eta$, see Proposition 5.4. Note that the nonvanishing of all $w_s$ implies that
$w_s\ne\eta$, see \tht{3.9}, and $p_s\ne 0$, see \tht{3.10}. Then $q_sr_s=-p_s^2\ne 0$, and $a_s\ne a_{s+1}$, $b_s\ne b_{s+1}$ by \tht{3.7}. 

Set $R_s=K_s(1-K_s)^{-1}$, where $K_s$ is the operator in $\ell^2(\{s,s+1,\dots\})$ defined by the discrete Bessel kernel restricted to $\{s,s+1,\dots\}$.\footnote{The existence of $(1-K_s)^{-1}$ follows, for example, from the fact that $D_s=\det(1-K_s)\ne 0$.} Then
$$
1+R_s(s,s)=\frac{\det(1-K_{s+1})}{\det(1-K_s)}=\frac{D_{s+1}}{D_s}\,.
\tag 4.1
$$

\proclaim{Proposition 4.1} Assume that $w_s\ne 0$ for $s\in\Z'_+$, and define $v_s$ as in Proposition 3.4. Then for any $s\in\Z'_+$ we have
$$
\left({\dfrac{D_{s+2}}{D_{s+1}}-\dfrac{D_{s+3}}{D_{s+2}}}\right)
=
\frac{(1-v_s^2)v_{s+1}^2}{v_{s}^2}\left({\dfrac{D_{s+1}}{D_s}-\dfrac{D_{s+2}}{D_{s+1}}}\right)\,.
\tag 4.2
$$ 
\endproclaim
\demo{Proof} By Theorem 2.3(ii) we have
$$
\gather
R_s(s,s)=g^t(s)m_s^{-1}(s)m_s'(s)f(s),\tag 4.3
\\ 
R_{s+1}(s+1,s+1)=g^t(s+1)m_{s+1}^{-1}(s+1)m_{s+1}'(s+1)f(s+1),\tag 4.4
\endgather
$$
where 
$$
f(x)=\left(\tfrac {\eta^x}{\Gamma(x+\frac 12)},0\right)^t\,,\quad
g(x)=\left(0,\,\tfrac {\eta^x}{\Gamma(x+\frac 12)}\right)^t.
$$

Let us evaluate \tht{4.3} using \tht{3.5}. We have
$$
\gather
m_s^{-1}(s)=\bmatrix \eta^{-1}(s+\frac 12)&0\\0& \eta(s+\frac 12)
 \endbmatrix m_{s+1}^{-1}(s+1)\bmatrix 0&-a_s\\b_s&
\eta^{-1}(s+\frac 12-p_s)
\endbmatrix,
\\
m_s'(s)=\frac{d}{d\ze}\left(\bmatrix \eta^{-1}(\ze-\frac 12-p_s)&a_s\\-b_s&0
\endbmatrix m_{s+1}(\ze)\bmatrix \eta(\ze-\frac 12)^{-1}&0\\0& \eta^{-1}(\ze-\frac 12) \endbmatrix\right)\Biggl|_{\ze=s+1}.
\endgather
$$

It is immediately verified that if the derivative in the last formula
falls on the third (diagonal) factor then the corresponding contribution
to \tht{4.3} vanishes. If the derivative falls on the second factor, 
the contribution to \tht{4.3} equals $R_{s+1}(s+1,s+1)$ because of \tht{4.4} and the equalities
$$
\gathered
g^t(s)\bmatrix \eta^{-1}(s+\frac 12)&0\\0& \eta(s+\frac 12)\endbmatrix
=g^t(s+1),\\ 
\bmatrix \eta(\ze-\frac 12)^{-1}&0\\0& \eta^{-1}(\ze-\frac 12) \endbmatrix f(s)=f(s+1).
\endgathered
$$

Finally, when the derivative falls on the first factor, using the relations above, we see that the contribution to \tht{4.3} is equal to
$$
g^t(s+1)m^{-1}_{s+1}(s+1)\bmatrix 0&0\\ \eta^{-1}b_s&0\endbmatrix
m_{s+1}(s+1)f(s+1).
\tag 4.5
$$

For $x\in\Z'_+$ denote
$$
m_x(x)=\bmatrix m^{11}_x&m^{12}_x\\m_x^{21}&m^{22}_x\endbmatrix.
$$
Since $\det m_x\equiv 1$, we have
$$
m_x(x)=\bmatrix m^{22}_x&-m^{12}_x\\-m_x^{21}&m^{11}_x\endbmatrix,
$$
and \tht{4.5} turns into
$$
\multline
g^t(s+1)\bmatrix m^{22}_{s+1}&-m^{12}_{s+1}\\-m_{s+1}^{21}&m^{11}_{s+1}
\endbmatrix\bmatrix 0&0\\ \eta^{-1}b_s&0\endbmatrix
\bmatrix m^{11}_{s+1}&m^{12}_{s+1}\\m_{s+1}^{21}&m^{22}_{s+1}\endbmatrix
f(s+1)\\ =\frac{\eta^{2s+1}}{\Gamma^2\left(s+\frac 32\right)}\,
b_s (m_{s+1}^{11})^2.
\endmultline
$$

Denote $R_s(s,s)-R_{s+1}(s+1,s+1)$ by $\delta_s$.
We have just proved that
$$
\delta_s=\frac{\eta^{2s+1}}{\Gamma^2\left(s+\frac 32\right)}\,
b_s (m_{s+1}^{11})^2.
\tag 4.6
$$

The (2,1) element of the equation \tht{3.5} at the point $\ze=s+1$ gives
$$
m_s^{21}=-b_sm_{s+1}^{11}\eta\, (s+\tfrac 12)^{-1}.
$$
Hence, $m_{s+1}^{11}=-b_s^{-1}m_s^{21}\eta^{-1}(s+\tfrac 12).$
Substituting this into the right--hand side of \tht{4.6} and using $a_s=b_s^{-1}$, we obtain
$$
\delta_s=\frac{\eta^{2s-1}}{\Gamma^2\left(s+\frac 12\right)}\,
a_s (m_{s}^{21})^2.
\tag 4.7
$$

Look at the residue of $m_{s+1}(\ze)$ at the point $\ze=s$. Since the jump matrix $-f(s)g^t(s)$ has zero first column, the residue itself also has zero first column. On the other hand, \tht{3.4} implies that
this residue equals $A_sm_s(s)$. Equating the (1,1) element of this matrix to zero we obtain
$$
p_s m_s^{11}=-q_sm_s^{21}
$$

By our assumption $q_s\ne 0$, hence, $m_s^{21}=-q_s^{-1}p_sm_s^{11}$.
Then \tht{4.7} turns into
$$
\delta_s=\frac{\eta^{2s-1}}{\Gamma^2\left(s+\frac 12\right)}\,
\frac {a_sp_s^2}{q_s^2}\, (m_{s}^{11})^2=-
\frac{\eta^{2s-1}}{\Gamma^2\left(s+\frac 12\right)}\,
\frac{w_s}{q_s}\, (m_{s}^{11})^2,
$$
where we used $p_s^2=-q_sr_s$ and $a_sr_s=w_s$. Comparing the last equality with \tht{4.6} we conclude that
$$
\delta_{s+1}=\frac{b_{s+1}w_{s+1}}{b_sw_s}\,\delta_{s}.
\tag 4.8
$$
The relation \tht{3.7} implies that $b_{s+1}b_s^{-1}=1-\eta^{-1}w_s=1-v_s^2$,
and \tht{4.2} follows. \qed
\enddemo

\proclaim{Corollary 4.2} Under the assumption of Proposition 4.1, there exist constants $\varkappa$ and $\nu$ such that
$$
\frac{D_{s+1}}{D_s}=\nu+\varkappa\cdot\eta b_s
\tag 4.9
$$
for all $s\in\Z'_+$.
\endproclaim
\demo{Proof}
By \tht{4.8} we have $\delta_s=\varkappa b_s w_s=\varkappa r_s$ for some
constant $\varkappa$. Then the second formula of \tht{3.7} implies
$$
\delta_s=\frac{D_{s+1}}{D_s}-\frac{D_{s+2}}{D_{s+1}}=\varkappa\cdot
\eta(b_s-b_{s+1}),
$$
and \tht{4.9} follows.\qed 
\enddemo

\head 5. Initial conditions for dPII
\endhead

The goal of this section is to provide initial conditions for the 
dPII in Proposition 3.4 and to find the constants $\varkappa$ and $\nu$
from Corollary 4.2. We will also prove that the assumption of Propositions 3.4 and 4.1 holds for generic $\eta$.

\proclaim{Lemma 5.1} The function $m_\frac 12(\ze)$ has the form
$$
m_{\frac 12}(\ze)=\bmatrix 1&0\\ -\sum\limits_{x\in\Z'_-}
\frac{\eta^{-2x}}{\Gamma^2(-x+\frac 12)}\,\dfrac 1{\ze-x}&1\endbmatrix.
\tag 5.1
$$
\endproclaim

\demo{Proof} A direct computation shows that the right--hand side of
\tht{5.1} solves the normalized DRHP $(\Z'_-,-fg^t)$. The uniqueness statement in Theorem 2.3(i) concludes the proof. \qed
\enddemo

\proclaim{Proposition 5.2} We have
$$
\gathered
A_\frac 12=\bmatrix p_\frac 12&q_\frac 12\\r_\frac 12&-p_\frac 12
\endbmatrix=\frac\eta {I_0(2\eta)}\bmatrix -I_1(2\eta)&-1\\
I_1^2(2\eta)& I_1(2\eta)\endbmatrix,\\
a_\frac 12=I_0^{-1}(2\eta),\quad b_\frac 12=I_0(2\eta),\\
w_\frac 12=a_\frac 12 r_\frac 12=\frac{\eta I_1^2(2\eta)}{I_0^2(2\eta)}\,,
\endgathered
\tag 5.2
$$
where $I_\nu(\,\cdot \,)$ is the $I$--Bessel function.
\endproclaim
\demo{Proof}
By \tht{3.4} we have
$$
m_\frac 32(\ze)=\left(I+\frac {A_\frac 12}{\ze-\frac 12}\right) m_\frac 12(\ze).
$$
Using \tht{5.1} and \tht{3.3} 
we see that the residue condition for $m_\frac 32(\ze)$
at the point $\ze=\frac 12$ looks as follows
$$
\gathered
A_\frac 12 \bmatrix 1&0\\ -\sum\limits_{x\in\Z'_-}
\frac{\eta^{-2x}}{(-x+\frac 12)\Gamma^2(-x+\frac 12)}&1\endbmatrix
\\
=\lim_{\ze\to\frac 12}\left(I+\frac {A_\frac 12}{\ze-\frac 12}\right)
\bmatrix 1&0 \\-\sum\limits_{x\in\Z'_-}
\frac{\eta^{-2x}}{\Gamma^2(-x+\frac 12)}\,\frac 1{\ze-x}&1\endbmatrix
\bmatrix 0& -\eta\\0&0\endbmatrix.
\endgathered
\tag 5.3
$$
Recall that
$$
I_0(2\eta)=\sum_{k\ge 0}\frac {\eta^{2k}}{\Gamma^2(k+1)}\,,\quad
I_1(2\eta)=\sum_{k\ge 0}\frac {\eta^{2k+1}}{(k+1)\Gamma^2(k+1)}\,.
$$
Since the (1,1) element of the right--hand side of \tht{5.3} vanishes, we obtain
$$
p_\frac 12=I_1(2\eta)\cdot q_\frac 12.
$$
Since the matrix $A_\frac 12$ is nilpotent, this implies that
$$
A_\frac 12=c\cdot \bmatrix -I_1(2\eta)&-1\\
I_1^2(2\eta)& I_1(2\eta)\endbmatrix
$$
with some constant $c$. Then the (1,2) element of \tht{5.3} gives
$$
-c=-\eta+c\cdot \sum_{x\in\Z'_-}\frac{\eta^{-2x+1}}
{(x+\frac 12)^2\,\Gamma^2(x+\frac 12)}\,.
$$ 
Hence, $c=\eta/I_0(2\eta)$, which proves the first line of \tht{5.2}.

Next, \tht{3.5} implies that $-b_\frac 12$ is the limit value of the (2,1) element of the matrix
$$
m_{\frac 12}(\ze-1)\bmatrix \eta^{-1}(\ze-\frac 12)&0\\
0&\eta(\ze-\frac 12)^{-1}\endbmatrix
$$
as $\ze\to\infty$. (Indeed, $m_{\frac 32}\to I$ as $\ze\to\infty$.) 
Since the (2,1) element of $m_\frac 12(\ze)$, see \tht{5.1}, has the form
$$
-\eta I_0(2\eta)\cdot \ze^{-1}+O(\ze^{-2}),\qquad \ze\to\infty,
$$
we conclude that $b_\frac 12=I_0(2\eta)$. This means that 
$a_\frac 12=b^{-1}_\frac 12=I_0^{-1}(2\eta)$, and the proof of Proposition 5.2 is complete.\qed
\enddemo

\proclaim{Corollary 5.3 (Initial conditions for dPII)} Assume that
$w_s\ne 0$ for all $s\in\Z'_+$. Then
the sequence $v_s$ from Proposition 3.4 can be defined by the initial conditions
$$
v_{-\frac 12}=\mp 1, \quad v_\frac 12=\pm\frac{I_1(2\eta)}{I_0(2\eta)}
\tag 5.4
$$
and the dPII equation \tht{3.13} with $s\ge \frac 12$.
\endproclaim
\demo{Proof} We will omit the argument $2\eta$ in the Bessel functions below. From \tht{5.2} and \tht{3.9} we obtain
$$
p_\frac 32=\frac{w_\frac 12}{w_\frac 12-\eta}-p_\frac 12=
\frac{I_1^2}{I_1^2-I_0^2}+\frac {\eta I_1}{I_0}\,,
$$
and by definition of $v_s$
$$
v_{\frac 12}=\pm \frac{I_1}{I_0}\,,\qquad v_\frac 32=
\frac{p_\frac 32}{\eta v_\frac 12}=\pm\frac{I_0I_1}{\eta(I_1^2-I_0^2)}\pm 1.
$$

This is exactly the same value of $p_\frac 32$ as we would obtain by 
substituting \tht{5.4} into \tht{3.13} with $s=\frac 12$. \qed
\enddemo

\proclaim{Proposition 5.4} The nonvanishing $w_s\ne 0$ for all $s\in\Z'_+$
holds for all but countably many $\eta$.
\endproclaim 
\demo{Proof} By Proposition 3.4, it is enough to show that for any $s\in\Z'_+$, $v_s$ defined via
\tht{5.4} and \tht{3.13} does not vanish for all but countably many 
values of $\eta$. Clearly, $v_s$ is a meromorphic function in $\eta$. 
Thus, it suffices to prove that $v_s$ does not vanish identically. 

It is easy to verify that the formulas
$$
\aligned
v_{2k-\frac 12}=& (-1)^k+\frac{kv}{v^2-1}\,\eta^{-1} +O(\eta^{-2}),\\
v_{2k+\frac 12}=&(-1)^k v^{(-1)^k} +O(\eta^{-1}),
\endaligned
\tag 5.5
$$
for $k=0,1,\dots$, define an asymptotic solution of \tht{3.13} as $\eta\to\infty$. Here $v$ is an arbitrary constant not equal to 0 or 1, and for a fixed $k$ the asymptotics is uniform in $v$ varying in any compact subset
of $\C\setminus\{0,1\}$. 

Since $I_1(2\eta)/I_0(2\eta)$ has zeros and poles with arbitrarily large $|\eta|$, see e.g. \cite{Er, 7.9}, we can find a sequence $\{\eta_m\}_{m=1}^\infty$  such that $|\eta_m|>m$ and  $|I_1(2\eta_m)/I_0(2\eta_m)|\in [\frac 13,\frac 23]$. Denote by $\{v_s^{(m)}\}_{s\in\Z_+'}$ the sequence defined by \tht{3.13} and \tht{5.4} with $\eta=\eta_m$. Then by \tht{5.5}, for any fixed $s\in\Z'_+$, $v_s^{(m)}$ is equal to 1, $-1$, $v$, or $-v^{-1}$,
where $v=I_1(2\eta_m)/I_0(2\eta_m)$, up to a correction term with goes to zero as $m\to\infty$. Hence, for large enough $m$ we have $v_s^{(m)}\ne 0$.
\qed
\enddemo

\proclaim{Corollary 5.5 (cf. Corollary 4.2)} For all but countably many values of $\eta$, we have $D_{s+1}/D_s=b_s$ for all $s\in\Z'_+$.
\endproclaim
\demo{Proof} Corollary 4.2 implies that $\delta_\frac 12=\varkappa r_\frac 12$.
The interpretation of $D_s$ as a Toeplitz determinant, see \tht{3.2}, easily implies that $\delta_{\frac 12}=I_1^2(2\eta)/I_0(2\eta)$. Comparing this with the value of $r_\frac 12$ from \tht{5.2} we see that $\varkappa=\eta^{-1}$. Furthermore, \tht{3.2} implies that
$D_{\frac 32}/D_\frac 12=I_0(2\eta)$. Since $b_\frac 12=I_0(2\eta)$,
we conclude that $\nu$ in \tht{4.9} vanishes.\qed
\enddemo

We can now state our final result.

\proclaim{Theorem 5.6} Let $s\in\Z'_+$, and let $D_s$ be the Fredholm
determinant of the discrete Bessel kernel as defined in the beginning of \S3. Define a sequence $\{v_s\}_{s\in\Z'_+}$ by initial conditions \tht{5.4} and the recurrence equation \tht{3.13}. Then for all but countably many values of $\eta$ we have
$$
v_s^2=1-\frac{D_sD_{s+2}}{D_{s+1}^2}\,.
\tag 5.6
$$
\endproclaim
\demo{Proof} Follows from Corollary 5.5, the definition of $\{v_s\}$ and 
the relation $\eta^{-1}w_s=1-b_{s+1}/b_s$, see \tht{3.7}.\qed
\enddemo

\example{Remark 5.7} If we extend $D_s$ to an entire function in $\eta$
using the Toeplitz determinant interpretation, see \tht{3.2}, then 
\tht{5.6} can be viewed as an equality of meromorphic functions.
\endexample

\head 6. Discrete ${}_2F_1$ kernel and dPV 
\endhead

In this section we obtain results similar to those of \S3, but for a more complicated kernel. In particular, we will see that the Fredholm 
determinant can be expressed through a solution of a dPV equation.

The discrete ${}_2F_1$ kernel plays a key role in harmonic analysis
on the infinite symmetric group. We refer the reader to \cite{BO2} and
\cite{BO3} for a detailed description. 

Set
$$
m(\ze)=\bmatrix F\left(-z,-z';\,\ze+\tfrac12;\,\tfrac \xi{\xi-1}\right)&
\frac {\sqrt{zz'\xi}}{1-\xi}\,\frac{F\left(1+z,1+z';\,-\ze+\tfrac 32;\,\tfrac  \xi{\xi-1}\right)}{-\ze+\tfrac 12}\\
-\frac {\sqrt{zz'\xi}}{1-\xi}\,\frac{F\left(1-z,1-z';\,\ze+\tfrac 32;\,\tfrac  \xi{\xi-1}\right)}{\ze+\tfrac 12}&F\left(z,z';\,-\ze+\tfrac12;\,\tfrac \xi{\xi-1}\right)\endbmatrix
\tag 6.1
$$
$$
\gathered
h_+(x)=\frac{(zz')^\frac 14\xi^{\frac x 2}(1-\xi)^{\frac{z+z'}2}\sqrt{(z+1)_{x-\frac 12}(z'+1)_{x-\frac 12}}}{\Gamma(x+\frac 12)}\,,\\
h_{-}(x)=\frac{(zz')^\frac 14\xi^{-\frac x 2}(1-\xi)^{-\frac{z+z'}2}\sqrt{(-z+1)_{-x-\frac 12}(-z'+1)_{-x-\frac 12}}}{\Gamma(-x+\frac 12)}\,.
\endgathered
$$
Here $F(a,b;c;u)$ is the Gauss hypergeometric function, $(a)_k=\Ga(a+k)/\Ga(a)$ is the Pochhammer symbol, $(z,z')$ are two complex parameters such that 
$(z+k)(z'+k)>0$ for all $k\in\Z$ (for instance, $z'=\bar{z}\in\C\setminus\Z$), $\xi\in(0,1)$ is also a parameter.
It is also convenient to assume that $z\ne z'$. 

The discrete ${}_2F_1$ kernel is a kernel on $\Z'\times \Z'$. For $x,y\in\Z'_+$ it  is defined by the formula
$$
K(x,y)=\cases
h_+(x)h_+(y)\,\dfrac{m^{21}(x)m^{11}(y)-m^{11}(x)m^{21}(y)}
{x-y}\,,& x\ne y\\
h_+^2(x) \left(\dfrac{d m^{21}(x)}{d x} m^{11}(x)-\dfrac {d m^{11}(x)}{d x} m^{21}(x)\right), &x=y.
\endcases
$$
The general definition for $x$ and $y$ not necessarily positive 
can be found in \cite{BO2}. Note that when one of the parameters $z,z'$
tends to an integer, the kernel (or rather the part of the kernel written above) turns into a Christoffel--Darboux kernel for Meixner orthogonal polynomials,
see \cite{BO2, \S4}.

As in \S3, for any $s\in \Z'_+$, denote by $K_s$ the operator in $\ell^2(\{s,s+1,\dots\})$ defined by the restriction of $K$ to
$\{s,s+1,\dots\}\times \{s,s+1,\dots\}$. Then $K_s$ is a positive trace class operator. We will be interested in the Fredholm determinants
$$
D_s=\det(1-K_s).
$$

These Fredholm determinants can be also expressed as Toeplitz determinants, see \cite{BOk, \S4} and also \cite{BW}, \cite{Bot}:
$$
D_s=(1-\xi)^{zz'}\cdot \det[t_{i-j}]_{i,j=1,\dots,s-\frac 12}\,
,
$$
where
$$ 
t_k=\cases \dfrac{\xi^{k/2}\Ga(-z+k)}{\Ga(-z)\Ga(k+1)}\,F(-z+k,-z';k+1;\xi),& k\ge 0,\\
\dfrac{\xi^{-k/2}\Ga(-z'-k)}{\Ga(-z')\Ga(-k+1)}\,F(-z'-k,-z;-k+1;\xi),& k<0.
\endcases
$$
The symbol of the Toeplitz determinants  is equal to
$$
\sum_{k\in\Z}t_k\ze^k=(1+\sqrt{\xi}\,\ze)^z\,(1+\sqrt{\xi}\,\ze^{-1})^{z'}.
$$

Note that in the Introduction we denoted $D_s$ by $q_{s-\frac 12}^{(z,z',\xi)}$.

The discrete ${}_2F_1$ kernel fits in Situation 2.1 of \S2: There exists an integrable kernel $L(x,y)$ on $\Z'\times \Z'$ such that it defines a bounded operator in $\ell^2(\Z')$, and $K=L(1+L)^{-1}$, see \cite{BO2}. Specifically, with respect to the splitting $\Z'=\Z'_+\sqcup \Z'_-$ the kernel $L$ has the form
$$
L(x,y)=\bmatrix 0& \frac{h_+(x)h_-(y)}{x-y}\\ \frac{h_-(x)h_+(y)}{x-y}&0\endbmatrix
$$
with $h_\pm$ as above. In the notation of \S2 we have $\x=\Z'$ and the matrix $m_\x(\ze)$ coincides with $m(\ze)$ of \tht{6.1}. Furthermore, 
$$
f_1(x)=g_2(x)=\cases h_+(x),&x\in\Z'_+,\\
0,&x\in\ \Z'_-,\endcases\quad
f_2(x)=g_1(x)=\cases 0,&x\in\Z'_+,\\
h_-(x),&x\in\Z'_-,\endcases
$$
and the jump matrix $w_\x(x)=-f(x)g(x)^t$ has the form
$$
w_\x(x)=\cases \bmatrix 0&-h_+^2(x)\\0&0\endbmatrix,&x\in\Z'_+,\\
\bmatrix 0&0\\-h_-^2(x)&0\endbmatrix,&x\in\Z'_-.
\endcases
$$
See \cite{Bor2} for details. 

Our next goal is to study the unique solution $m_\z$ of the normalized DRHP $(\z,w_\x)$, where 
$$
\z=\x\setminus\y=\Z\setminus\{s,s+1,\dots\}=\{\dots,s-2,s-1\},
$$
see Theorem 2.3(i). As in \S3, we redenote $m_\z$ by $m_s$. 
We also denote $\z_s=\{\dots,s-2,s-1\}$.
Recall that $\det m_s\equiv 1$. Set
$$
\Xi=\bmatrix \xi^\frac 12&0\\0 &\xi^{-\frac 12}\endbmatrix.
$$

The following statement is an analog of Proposition 3.2. 
\proclaim{Proposition 6.1 (Lax pair)} For any $s\in\Z'_+$ there exist
constant $2\times 2$ matrices $A_s$ and $B_s$ such that
$$
\gather
m_{s+1}(\ze)=\left(I+\frac {A_s}{\ze-s}\right)m_s(\ze), \tag 6.2\\
m_s(\ze-1)=\Xi^{-1}
\left(I+\frac {B_s}{\ze+z-\frac 12}\right) m_{s+1}(\ze)\bmatrix \frac{\ze+z-\frac 12}{\ze-\frac 12}&0\\0& \frac{\ze-\frac 12}{\ze+z'-\frac 12} 
\endbmatrix\Xi\,.
\tag 6.3
\endgather
$$
Furthermore, 
$$
\gathered
\operatorname{Tr} A_s=\operatorname{Tr} B_s=0,\qquad
\det A_s=0,\quad \det B_s=z'-z,\\ 
A_s+B_s=
\bmatrix -z& *\\*&z'\endbmatrix.
\endgathered
\tag 6.4
$$
\endproclaim
\demo{Proof} The proofs of \tht{6.2} and of the fact that $A_s$ is nilpotent are very similar to the proofs of analogous statements 
in Proposition 3.2. 

Similarly to the proof of Proposition 3.2, it is easy to show that
$$
m_s(\ze-1)\, \Xi^{-1}
\bmatrix \frac{\ze-\frac 12}{\ze+z-\frac 12}&0\\0& \frac{\ze+z'-\frac 12}{\ze-\frac 12}\endbmatrix m_{s+1}^{-1}(\ze)
\tag 6.5
$$
has no singularities in $\Z'$. Thus, the only possible
singularity of \tht{6.5} is a simple pole at $\ze=-z+\frac 12$. (Note that by our assumption on the parameters $z,z'$, they are both nonintegral, and $-z+\frac 12\notin\Z'$.) Denote the residue by $C_s$. Since both $m_s(\ze)$ and $m_{s+1}(\ze)$ tend to $I$ as $\ze\to\infty$, we see that \tht{6.5}
tends to $\Xi^{-1}$ as $\ze\to\infty$, and by Liouville's theorem we
obtain \tht{6.3} with $B_s=\Xi C_s$.

Taking the determinants of both sides of \tht{6.3} we see that
$$
\det\left(I+\frac{B_s}{\ze+z-\frac 12}\right)=1+\frac{z'-z}{\ze+z-\frac 12}\,.
$$
This implies that $\operatorname{Tr} B_s=z'-z$ and $\det B_s=0$.

Finally, let us substitute \tht{6.2} into the right--hand side of  \tht{6.3}. We obtain
$$
\gathered
m_s(\ze-1)=\Xi^{-1}
\left(I+\frac {B_s}{\ze+z-\frac 12}\right) \left(I+\frac {A_s}{\ze-s}\right)m_s(\ze) \bmatrix \frac{\ze+z-\frac 12}{\ze-\frac 12}&0\\0& \frac{\ze-\frac 12}{\ze+z'-\frac 12} 
\endbmatrix\Xi\,.
\endgathered
\tag 6.6
$$

We know that $m_s(\ze)=I+m_s^{(1)}\ze^{-1}+O(\ze^{-2})$ as $\ze\to\infty$, with a constant matrix $m_s^{(1)}$. (This follows from the general formula \cite{Bor2, (4.9)}.) Substituting this asymptotic relation into both sides of \tht{6.6} and comparing the asymptotics of the diagonal elements of both sides at $\ze=\infty$ we obtain 
$$
A_s^{11}+B_s^{11}=-z,\quad A_s^{22}+B_s^{22}=z'. \qed
$$
\enddemo

For any $2\times 2$ matrix $M$ we will denote by $M^\Xi$ the matrix 
$\Xi M \Xi^{-1}$.

\proclaim{Corollary 6.2 (Consistency relations)} For any $s\in\Z'_+$ we have
$$
A_{s+1}-A_s^\Xi=B_s-B_{s+1}=\frac {A_s^\Xi B_s-B_{s+1}A_{s+1}}{s+z+\frac 12}\,.\tag 6.7
$$
\endproclaim
\demo{Proof} Similarly to the proof of Corollary 3.3, we compute $m_{s+1}(\ze)$ in two different ways. Using \tht{6.2} and then \tht{6.3}
we get
$$
m_{s+1}(\ze)=\left(I+\frac {A_s}{\ze-s}\right)\Xi^{-1}
\left(I+\frac {B_s}{\ze+z+\frac 12}\right)m_{s+1}(\ze+1)\bmatrix \frac{\ze+z+\frac 12}{\ze+\frac 12}&0\\0& \frac{\ze+\frac 12}{\ze+z'+\frac 12} 
\endbmatrix\Xi. 
$$
On the other hand, using \tht{6.3} first and then \tht{6.2} we get
$$
m_{s+1}(\ze)=\Xi^{-1}
\left(I+\frac {B_{s+1}}{\ze+z+\frac 12}\right)\left(I+\frac {A_{s+1}}{\ze-s}\right)m_{s+1}(\ze+1)\bmatrix \frac{\ze+z+\frac 12}{\ze+\frac 12}&0\\0& \frac{\ze+\frac 12}{\ze+z'+\frac 12} 
\endbmatrix\Xi. 
$$
Comparing the results we see that
$$
\left(I+\frac {A_s}{\ze-s}\right)\Xi^{-1}
\left(I+\frac {B_s}{\ze+z+\frac 12}\right)=\Xi^{-1}
\left(I+\frac {B_{s+1}}{\ze+z+\frac 12}\right)\left(I+\frac {A_{s+1}}{\ze-s}\right).
$$
Equating the residues of both sides at $\ze=s$ and $\ze=-z-\frac 12$ 
yields \tht{6.7}.\qed
\enddemo

In order to proceed to deriving the dPV equation we need to impose certain nondegeneracy conditions on $A_s,B_s$ similar to the condition $w_s\ne 0$ used in Proposition 3.4.

We will say that matrices $A_s$ and $B_s$ are {\it generic} if we can uniquely parameterize them by
$$
A_s=(z+b_s)\bmatrix -1&-\alpha_s\beta_s\\ 1/(\alpha_s\beta_s)&1\endbmatrix,\quad
B_s=\bmatrix b_s&b_s\beta_s\\ (z'-z-b_s)/\beta_s&z'-z-b_s\endbmatrix
\tag 6.8
$$
with some $b_s$, $\alpha_s\ne 0$, $\beta_s\ne 0$. This is certainly true
if the off--diagonal entries of $A_s$ and $B_s$ are nonzero:
$$
A_s^{12},\,A_s^{21},\,B_s^{12},\, B_s^{21}\ne 0.
$$
It is also true if
$$
A_s^{12},\,A_s^{21}, B_s^{21}, B_s^{22}\ne 0,\quad B_s^{11}=B_s^{12}=0.
$$
Then $b_s=0$, $\beta_s=(z'-z)/B_s^{21}$, $\alpha_s=-A_s^{12}/(z\beta_s)$. Recall that $z\ne z'$ by the assumption.
Similarly, if
$$
A_s^{12},\,A_s^{21}, B_s^{11}, B_s^{12}\ne 0,\quad B_s^{21}=B_s^{22}=0.
$$
Then $b_s=z'-z$, $\beta_s=B_s^{12}/(z'-z)$, $\alpha_s=-A_s^{12}/(z\beta_s)$. These three cases exhaust all possibilities.

\proclaim{Proposition 6.3 (dPV)} Fix $s\in\Z'_+$. Assume that the matrices $A_s$ and $B_s$ have nonzero off--diagonal entries, and in the notation \tht{6.8}
$$
\xi\alpha_s\ne 1,\quad b_s+\frac{z'+s+\tfrac 12}{1-\xi\alpha_s}\notin
\left\{-z,\,z'-z,\,z'+s+\tfrac 12,\,z'-z+s+\tfrac 12\right\}.
\tag 6.9
$$
Then the matrices $A_{s+1}$ and $B_{s+1}$ are generic, and 
$$
\gather
\alpha_{s+1}=\frac{\left(b_s+\dfrac{z'+s+\tfrac 12}{1-\xi\alpha_s}-\left(z'+s+\frac 12\right)\right)\left(b_s+\dfrac{z'+s+\tfrac 12}{1-\xi\alpha_s}-\left(z'-z+s+\frac 12\right)\right)}
{\xi\alpha_s\left(b_s+\dfrac{z'+s+\tfrac 12}{1-\xi\alpha_s}+z\right)\left(b_s+\dfrac{z'+s+\tfrac 12}{1-\xi\alpha_s}+z-z'\right)}\,,
\tag 6.10\\
b_{s+1}=-b_s-\frac{z'+s+\tfrac 12}{1-\xi\alpha_s}+\frac{z+s+\frac 12}{1-\alpha_{s+1}}-2z+z',
\tag 6.11\\
{\beta_{s+1}}=\frac {\xi\alpha_s\left(b_s+\dfrac{z'+s+\tfrac 12}{1-\xi\alpha_s}+z\right)}{b_s+\dfrac{z'+s+\tfrac 12}{1-\xi\alpha_s}-\left(z'-z+s+\frac 12\right)}\cdot{\beta_s}\,.
\tag 6.12
\endgather
$$
\endproclaim
\example{Remarks 6.4} 1. The conditions \tht{6.9} mean that
four factors in the right--hand side of \tht{6.10} as well as two factors in the right--hand side of \tht{6.12} and the denominator $(1-\xi\alpha_s)$ in \tht{6.11} do not vanish. As we will see in the proof, \tht{6.9} also implies that $\alpha_{s+1}\ne 1$, hence, the right--hand side of \tht{6.11} is also well--defined.

2. If we define $(\alpha_{s+1},\,\beta_{s+1},\,b_{s+1})$ via \tht{6.10}--\tht{6.12} and define the matrices $A_{s+1},\, B_{s+1}$ using \tht{6.8} (with $s$ replaced by $s+1$), then the consistency relations \tht{6.7} will hold. 

3. The relations \tht{6.10}, \tht{6.11} define a map $(\alpha_s,b_s)\mapsto (\alpha_{s+1}, b_{s+1})$. The inverse map has
a similar form. Let us assume that $A_{s+1}$ and $B_{s+1}$ have nonzero off--diagonal entries, and
$$
\alpha_{s+1}\ne 1,\quad b_{s+1}-\frac{z+s+\frac 12}{1-\alpha_{s+1}}\notin
\left\{-z,z'-z,-(z+s+\tfrac 12),-(2z+s+\tfrac 12)\right\}.
$$
Then the matrices $A_s$ and $B_s$ are generic and
$$
\gathered
\alpha_{s}=\frac{
\left(b_{s+1}-\dfrac{z+s+\tfrac 12}{1-\alpha_{s+1}}+z+s+\frac 12\right)
\left(b_{s+1}-\dfrac{z+s+\tfrac 12}{1-\alpha_{s+1}}+2z+s+\frac 12\right)}
{
\xi\alpha_{s+1}\left(b_{s+1}-\dfrac{z+s+\tfrac 12}{1-\alpha_{s+1}}+z\right)\left(b_{s+1}-\dfrac{z+s+\tfrac 12}{1-\alpha_{s+1}}+z-z'\right)}\,,
\\
b_{s}=-b_{s+1}+\frac{z+s+\frac 12}{1-\alpha_{s+1}}-\frac{z'+s+\tfrac 12}{1-\xi\alpha_s}-2z+z',\\
{\beta_{s}}=\frac{\alpha_{s+1}\left(b_{s+1}-\dfrac{z+s+\tfrac 12}{1-\alpha_{s+1}}+z\right)}{b_{s+1}-\dfrac{z+s+\tfrac 12}{1-\alpha_{s+1}}+2z+s+\frac 12}\cdot {\beta_{s+1}}\,.
\endgathered
\tag 6.13
$$

4. Let us introduce a new variable 
$$
c_s=b_s+\frac{z'+s+\frac 12}{1-\xi\alpha_s}+z
$$ 
and rewrite \tht{6.9}--\tht{6.12} in using $(c_s,c_{s+1})$ instead of $(b_s, b_{s+1})$. Then the formulas become slightly simpler:
$$
\gathered
\xi\alpha_s\ne 1,\quad c_s\notin
\left\{0,\,z',\,z'+s+\tfrac 12,\,z+z'+s+\tfrac 12\right\},\\
\alpha_s\alpha_{s+1}=\frac{\left(c_s-\left(z+z'+s+\frac 12\right)\right)\left(c_s-\left(z'+s+\frac 12\right)\right)}
{\xi c_s\left(c_s-z'\right)}\,,\\
c_s+c_{s+1}=\frac{z+s+\frac 12}{1-\alpha_{s+1}}+\frac{z'+s+\frac 32}{1-\xi\alpha_{s+1}}+z',\\
\frac{\beta_{s+1}}{\beta_s}=\frac {\xi\alpha_sc_s}{c_s-
\left(z'+s+\frac 12\right)}\,.
\endgathered
\tag 6.14
$$

Similarly, if we set
$$
d_s=b_s-\frac{z+s-\frac12}{1-\alpha_s}+z
$$
then the nondegeneracy condition and \tht{6.13} take the form
$$
\gathered
\alpha_{s+1}\ne 1,\quad d_{s+1}\notin
\left\{0,\,z',\,-(s+\tfrac 12),\,-(z+s+\tfrac 12)\right\},\\
\alpha_{s}\alpha_{s+1}=\frac{
\left(d_{s+1}+s+\frac 12\right)
\left(d_{s+1}+z+s+\frac 12\right)}
{
\xi d_{s+1}\left(d_{s+1}-z'\right)}\,,
\\
d_{s}+
d_{s+1}=-\frac{z'+s+\tfrac 12}{1-\xi\alpha_s}-\frac{z+s-\frac 12}{1-\alpha_s}+z',
\\
\frac{\beta_{s}}{\beta_{s+1}}=\frac{\alpha_{s+1}d_{s+1}}{d_{s+1}+z+s+\frac 12}\,.
\endgathered
\tag 6.15
$$
The relations between $(\alpha_s,d_s)$ and $(\alpha_{s+1},d_{s+1})$ form a special case of the difference Painlev\'e V equation of 
\cite{Sak, \S7}. The parameters $a_0,\dots, a_4$ in our case 
are as follows:
$$
\gathered
a_0=-(z'+s+\tfrac 12),\quad a_1=-(z+s+\tfrac 12),\quad a_2=s+\tfrac 32, \\
a_3=z',\quad a_4=z,\quad \lambda=a_0+a_1+2a_2+a_3+a_4=1.
\endgathered
$$
\endexample

\demo{Proof of Proposition 6.3} 

\noindent {\bf Step 1.} Let us first show that the matrices $A_{s+1}$ and $B_{s+1}$ are generic. By \tht{6.4} we can write these matrices in the form
$$
A_{s+1}=\bmatrix -z-b & A^{12}\\A^{21} &z+b\endbmatrix,\quad
B_{s+1}=\bmatrix b & B^{12}\\ B^{21} & z'-z-b \endbmatrix
$$
with some constants $b,\, A^{12},\, A^{21},\, B^{12},\, B^{21}$ such that
$$
\det A_{s+1}=-(z+b)^2-A^{12}A^{21}=0,\quad \det B_{s+1}=b(z'-z-b)-B^{12}B^{21}=0.
\tag 6.16
$$
Let us prove that $A^{12},\, A^{21}$ are nonzero.

Assume that $A^{21}=0$. Then $b=-z$ from \tht{6.16}. 
The (1,1)--element of the second equality in \tht{6.7} gives
$$
b_s-b=\frac{-(z+b_s)b_s-\xi\alpha_s(z+b_s)(z'-z-b_s)+
b(z+b)-B^{12}A^{21}}{z+s+\frac 12}\,.
\tag 6.17
$$
Hence, 
$$
(z+b_s)\left(1+\frac{b_s+\xi\alpha_s(z'-z-b_s)}{z+s+\frac 12}\right)=0.
$$
Since $\det A_s=0$ and $A_s$ has nonzero off--diagonal entries, its diagonal elements are nonzero, and $z+b_s\ne 0$. Thus, the second factor in the last equality vanishes
and 
$$
b_s+\frac{z'+s+\frac 12}{1-\xi\alpha_s}=z'-z
$$
which contradicts \tht{6.9}.

Assume that $A^{12}=0$. Then, again, $b=-z$, and the (2,2)--element of the second equality in \tht{6.7} gives
$$
b-b_s=\frac{(z+b_s)(z'-z-b_s)+b_s(z+b_s)(\xi\alpha_s)^{-1}-(z+b)(z'-z-b)-B^{21}A^{12}}{z+s+\frac 12}\,.
\tag 6.18
$$ 
As above, this leads to 
$$
b_s+\frac{z'+s+\frac 12}{1-\xi\alpha_s}=z'+s+\tfrac 12
$$
which again contradicts \tht{6.9}.

Now let us proceed to $B_{s+1}$. Assume $B^{12}=0$. Then \tht{6.16} implies that either $b=0$ or $b=z'-z$. If $b=z'-z$ then similarly to the above \tht{6.17} yields 
$$
b_s+\frac{z'+s+\frac 12}{1-\xi\alpha_s}=-z
$$
which cannot happen by \tht{6.9}. (In this reduction one needs to know that $b_s+z-z'\ne 0$. This follows from the fact that $\det B_s=0$ and the hypothesis that the off--diagonal entries of $B_s$ are nonzero. It follows that the diagonal entries, including $b_s+z-z'$ are also nonzero.) If $b=0$ then the only situation when $B_{s+1}$ is not generic is when $B^{21}=0$, see below.

Finally, assume that $B^{21}=0$. Then by \tht{6.16} either $b=0$ or $b=z'-z$. If $b=z'-z$ and $B^{12}\ne 0$ then $B_{s+1}$ is generic, and the case $B^{12}=0$ was already considered above. If $b=0$ then \tht{6.18}
yields
$$
b_s+\frac{z'+s+\frac 12}{1-\xi\alpha_s}=z'-z+s+\tfrac 12\,.
$$ 
Once again, this contradicts \tht{6.9}. 

Thus, we have proved that $A_{s+1}$ and $B_{s+1}$ are generic.

\noindent {\bf Step 2.} From now on we will use the form \tht{6.8} for 
$A_{s+1}$ and $B_{s+1}$. Since $A_{s+1}$, $B_{s+1}$ are generic,  $\alpha_{s+1},\,b_{s+1},\,\beta_{s+1}$ are well--defined.
In this part of the proof we derive \tht{6.10}--\tht{6.12} under the additional assumption $b_{s+1}\ne b_s$.

Relations \tht{6.17} and \tht{6.18} rewritten in the notation \tht{6.8} give
$$
\gather
b_s-b_{s+1}=\frac{-(z+b_s)(b_s+\xi\alpha_s(z'-z-b_s))+
(z+b_{s+1})b_{s+1}(1-1/\alpha_{s+1})}{z+s+\frac 12}\,,
\tag 6.19\\
b_{s+1}-b_s=\frac{(z+b_s)((z'-z-b_s)+b_s/(\xi\alpha_s))-(z+b_{s+1})(z'-z-b_{s+1})(1-\alpha_{s+1})}{z+s+\frac 12}\,.
\tag 6.20
\endgather
$$
Add \tht{6.19} multiplied by $\alpha_{s+1}$ and \tht{6.20} multiplied by $\xi\alpha_s$. We obtain
$$
(b_{s+1}-b_s)\left(\xi\alpha_s-\alpha_{s+1}+\frac{(1-\alpha_{s+1})
((1-\xi\alpha_s)(b_{s+1}+b_{s}+2z-z')+z-z')}{s+z+\frac 12}\right)=0.
$$
We assumed that $b_{s+1}-b_s\ne 0$, and vanishing of the second factor 
is easily seen to be equivalent to \tht{6.11}. (Note that $\alpha_{s+1}$ cannot equal 1. Indeed, in that case vanishing of the second factor would imply $\xi\alpha_s=1$, which contradicts our hypothesis.) Substituting $b_{s+1}$ from \tht{6.11} into either \tht{6.19} or \tht{6.20} and solving for
$\alpha_{s+1}$ yields \tht{6.10}.

To prove \tht{6.12} we look at the (1,2) and (2,1)--elements of the first equality $A_s^\Xi+B_s=A_{s+1}+B_{s+1}$ in \tht{6.7}:
$$
\gathered
\beta_s(-\xi\alpha_s(z+b_s)+b_s)=\beta_{s+1}(
-\alpha_{s+1}(z+b_{s+1})+b_{s+1}),\\
\beta_s^{-1}((z+b_s)/(\xi\alpha_s)+z'-z-b_s)=
\beta_{s+1}^{-1}((z+b_{s+1})/\alpha_{s+1}+z'-z-b_{s+1}).
\endgathered
\tag 6.21
$$
Substituting $\alpha_{s+1}$ and $b_{s+1}$ from \tht{6.10}, \tht{6.11} we
obtain (after some tedious work)
$$
\gathered
(b_s-\xi\alpha_s(z+b_s))\left(\beta_s-\frac {b_s+\frac{z'+s+\tfrac 12}{1-\xi\alpha_s}-\left(z'-z+s+\frac 12\right)}{\xi\alpha_s\left(b_s+\frac{z'+s+\tfrac 12}{1-\xi\alpha_s}+z\right)}\,\beta_{s+1}\right)=0,\\
\left(\frac{z+b_s}{\xi\alpha_s}+z'-z-b_s\right)\left(\beta_s^{-1}-\frac {\xi\alpha_s\left(b_s+\frac{z'+s+\tfrac 12}{1-\xi\alpha_s}+z\right)}{b_s+\frac{z'+s+\tfrac 12}{1-\xi\alpha_s}-\left(z'-z+s+\frac 12\right)}\,\beta_{s+1}^{-1}\right)=0.
\endgathered
$$
This yields \tht{6.12} if at least one of the prefactors 
$$
b_s-\xi\alpha_s(z+b_s) \quad\text{  and  }\quad \frac{z+b_s}{\xi\alpha_s}+z'-z-b_s
\tag 6.22
$$
is nonzero. But if they are both zero then we obtain $\alpha_s=-z/(\xi z')$, $b_s=-z^2/(z+z')$. The evaluation of $b_{s+1}$ through \tht{6.11} gives $b_{s+1}=b_s$, which contradicts our assumption.

Thus, we have proved \tht{6.10}--\tht{6.12} under the assumption $b_{s+1}\ne b_s$.

\noindent {\bf Step 3.}
Let us assume that $b_{s+1}=b_s$. Recall that $b_s,\,b_s+z,\,b_s+z-z'$ are all nonzero because otherwise $A_s$ or $B_s$ would have zero off--diagonal entries which is impossible by the hypothesis. Equation \tht{6.19} implies
$$
\alpha_{s+1}=\frac{b_s}{\xi\alpha_s(b_s+z-z')}\,.
\tag 6.23
$$
Let us show that $\alpha_{s+1}\ne 1$. Using the relation $b_{s+1}=b_s$, \tht{6.23}, and $\alpha_{s+1}=1$ we obtain
$$
\gathered
(A_{s+1}+B_{s+1}-A_s^\Xi-B_s)^{12}=\frac{z'b_s\beta_s-z(b_s+z-z')\beta_{s+1}}{b_s+z-z'}\,,\\
\left(A_{s+1}-A_s^\Xi-\frac {A_s^\Xi B_s-B_{s+1}A_{s+1}}{s+z+\frac 12}\right)^{12}=\frac{(z+b_s)(b_s\beta_s-(b_s+z-z')\beta_{s+1})}{b_s+z-z'}\,.
\endgathered
$$
Since $z\ne z'$, both these expressions cannot vanish simultaneously, which contradicts \tht{6.7}. Thus, $\alpha_{s+1}\ne 1$.

In order to proceed we assume that both expressions \tht{6.22} are nonzero. The case when one of them vanishes will be considered in Step 4 below. 

The first relation \tht{6.21} implies 
$$
\beta_{s+1}=\frac{b_s-\xi\alpha_s(z+b_s)}{b_{s+1}-\alpha_{s+1}(z+b_{s+1})}\,\beta_s.
\tag 6.24
$$

Let us denote by $\alpha_{s+1}^o$ and $b_{s+1}^o$ the differences of the left and right--hand sides of \tht{6.10} and \tht{6.11}. Then with the substitutions $b_{s+1}=b_s$, \tht{6.23}, \tht{6.24} one computes that
$$
\gathered
\frac{\beta_s(b_s+z)(b_s+z-z')(1-\xi\alpha_s)^2
\left(b_s+\frac{z'+s+\tfrac 12}{1-\xi\alpha_s}+z\right)
\left(b_s+\frac{z'+s+\tfrac 12}{1-\xi\alpha_s}+z-z'\right)}{
(z+s+\frac 12)(z'+s+\frac 12)\left(\frac{z+b_s}{\xi\alpha_s}+z'-z-b_s\right)}\cdot\alpha_{s+1}^o\\
=\left(A_{s+1}-A_s^\Xi-\frac {A_s^\Xi B_s-B_{s+1}A_{s+1}}{s+z+\frac 12}\right)^{12}\\=
 \frac{(1-\xi\alpha_s)\beta_s(z+b_s)(b_s-\xi\alpha_s(b_s+z-z'))}
{\xi\alpha_s\left(\frac{z+b_s}{\xi\alpha_s}+z'-z-b_s\right)}\cdot b_{s+1}^o\,.
\endgathered
$$
Thanks to all the assumptions, none of the prefactors of $\alpha_{s+1}^o$ and $b_{s+1}^o$ vanishes. Thus, \tht{6.7} implies
\tht{6.10}, \tht{6.11}. Then \tht{6.12} follows in the same way as in Step 2 above. 

\noindent {\bf Step 4.} Finally, assume that $b_{s+1}=b_s$ and one of the expressions \tht{6.22} vanishes. 
Let us show that the other one also
vanishes. Indeed, if $b_s-\xi\alpha_s(z+b_s)=0$ then the first
equation \tht{6.21} implies that $b_{s+1}-\alpha_{s+1}(z+b_{s+1})=0$.
Substituting $b_{s+1}=b_s$ and \tht{6.23} we get
$$
b_{s+1}-\alpha_{s+1}(z+b_{s+1})=\frac{b_s\left( (z+b_s)/(\xi\alpha_s)+z'-z-b_s\right)}{b_s+z-z'}=0,
$$
hence, $(z+b_s)/(\xi\alpha_s)+z'-z-b_s=0.$  Conversely, if $(z+b_s)/(\xi\alpha_s)+z'-z-b_s=0$ then the second equation of \tht{6.21} implies $(z+b_{s+1})/\alpha_{s+1}+z'-z-b_{s+1}=0$. With the same substitutions we obtain
$$
(z+b_{s+1})/\alpha_{s+1}+z'-z-b_{s+1}=(\xi\alpha_s(z+b_s)-b_s)/(\xi\alpha_s)=0,
$$
hence, $b_s-\xi\alpha_s(z+b_s)=0$. 

On the other hand, if both expressions \tht{6.22} vanish, we obtain
$$
\alpha_s=-\frac z{\xi z'}\,, \quad b_s=-\frac{z^2}{z+z'}\,.
\tag 6.25
$$
In particular, this means that $z+z'\ne 0$. 
Hence, using \tht{6.23}, we get
$$
\alpha_{s+1}=-\frac z{z'}\,,\quad b_{s+1}=b_s=-\frac{z^2}{z+z'}\,.
\tag 6.26
$$
But \tht{6.26} exactly is exactly what one gets by substituting \tht{6.25} into \tht{6.10}, \tht{6.11}. To prove \tht{6.12} we compute, using \tht{6.25},
$$
\left(A_{s+1}-A_s^\Xi-\frac {A_s^\Xi B_s-B_{s+1}A_{s+1}}{s+z+\frac 12}\right)^{12}=\frac{z^2 (\beta_{s+1}(s+\frac 12)-\beta_s(z+z'+s+\frac 12))}{(z+z')(z+s+\frac 12)}\,.
$$
By \tht{6.7} this implies that $\beta_{s+1}=\beta_s\cdot(z+z'+s+\frac 12)/(s+\frac 12)$ which coincides with \tht{6.12} under the conditions \tht{6.25}. The proof of Proposition 6.3 is complete. \qed
\enddemo

\head 7. Fredholm determinant and dPV
\endhead

Recall that in the beginning of the previous section we introduced $D_s$ as the Fredholm determinant of $(1-K_s)$, where $K_s$ is the hypergeometric kernel restricted to $\{s,s+1,\dots\}$.
In this section we relate $D_s$ to the sequences $\{b_s\},\, \{\alpha_s\},\,\{\beta_s\}$. Our goal is to prove the following

\proclaim{Proposition 7.1} Fix $s\in\Z'_+$.
Under the assumptions of Proposition 6.3, we have
$$
\gather
\left(\frac{D_{s+1}}{D_s}-\frac{D_{s+2}}{D_{s+1}}\right)
((1-\xi\alpha_{s+1})(b_{s+1}+z-z')-z+z')\\=
\left(\frac{D_{s+2}}{D_{s+1}}-\frac{D_{s+3}}{D_{s+2}}\right)
((1-\xi\alpha_s)(b_s+z-z')-z+z')
\tag 7.1
\\ 
\times \frac{(z+s+\frac 32)(z'+s+\frac 32)\left(b_s+\dfrac{s+z'+\frac12}{1-\xi\alpha_s}-(z'-z+s+\frac 12)\right)}
{(1-\xi\alpha_{s+1})(1-\xi\alpha_s)\,\alpha_s
\left(b_s+\dfrac{s+z'+\frac12}{1-\xi\alpha_s}+z\right)\left(b_s+\dfrac{s+z'+\frac12}{1-\xi\alpha_s}+z-z'\right)^2}\,.
\endgather
$$
\endproclaim
\example{Remark 7.2} Using the variables $c_s$ of Remark 6.4(4) we can rewrite \tht{7.1} as
$$
\gathered
\left(\frac{D_{s+1}}{D_s}-\frac{D_{s+2}}{D_{s+1}}\right)
\left((1-\xi\alpha_{s+1})(c_{s+1}-z')-(z+s+\tfrac 32)\right)\\=
\left(\frac{D_{s+2}}{D_{s+1}}-\frac{D_{s+3}}{D_{s+2}}\right)
\left((1-\xi\alpha_{s})(c_{s}-z')-(z+s+\tfrac 12)\right)
\\ 
\times \frac{(z+s+\frac 32)(z'+s+\frac 32)\left(c_s-(z'+s+\frac 12)\right)}
{(1-\xi\alpha_{s+1})(1-\xi\alpha_s)\,\alpha_s
c_s\left(c_s-z'\right)^2}\,.
\endgathered
$$
\endexample

To prove Proposition 7.1 it suffices to prove the following two lemmas.
\proclaim{Lemma 7.3}
Assume that the matrices $A_s$ and $B_s$ have nonzero off--diagonal entries. Then
$$
\gathered
\frac{D_{s+1}}{D_s}-\frac{D_{s+2}}{D_{s+1}}=\frac{(zz')^\frac 12(1-\xi)^{z+z'} \xi^{s-1}(z+1)_{s-\frac 12}(z'+1)_{s-\frac 12}}{(z+s+\frac 12)(z'+s+\frac 12)
\Gamma^2(s+\frac 12)}\\ \times (1-\xi\alpha_s)
\left((1-\xi\alpha_{s})(b_{s}+z-z')-z+z'\right)\cdot \beta_s\gamma_s^2
\endgathered
\tag 7.2
$$
where $\gamma_s$ is the $(2,1)$--entry of the matrix $m_s(s)$.
\endproclaim

\proclaim{Lemma 7.4} Under the assumptions of Proposition 6.3 we have
$$
\gamma_{s+1}=\frac{(s+\frac 12)(1-\xi\alpha_s)
\left(b_s+\dfrac{s+z'+\frac12}{1-\xi\alpha_s}+z-z'\right)}
{\xi(z+s+\frac 12)(z'+s+\frac 12)}\cdot \gamma_s
\tag 7.3
$$
where $\gamma_s=m_s^{21}(s)$, $\gamma_{s+1}=m_{s+1}^{21}(s+1)$.
\endproclaim
The relation \tht{7.1} is a direct consequence of \tht{7.2}, \tht{7.3}, and \tht{6.12}. 

\demo{Proof of Lemma 7.3} The proof reminds that of Proposition 4.1 but is a little more technically involved. 
Theorem 2.3(ii) gives
$$
\gather
R_s(s,s)=g^t(s)m_s^{-1}(s)m_s'(s)f(s),\tag 7.4
\\ 
R_{s+1}(s+1,s+1)=g^t(s+1)m_{s+1}^{-1}(s+1)m_{s+1}'(s+1)f(s+1),\tag 7.5
\endgather
$$
where (see \S6 for the definition of $h_+$)
$$
f(x)=\left(h_+(x),0\right)^t\,,\quad
g(x)=\left(0,\,h_+(x)\right)^t.
$$
(Recall that $R_s=K_s(1-K_s)^{-1}$ and $R_s(s,s)=D_{s+1}/D_s-1$.)

Let us plug the expression for $m_{s+1}$ from \tht{6.3} into \tht{7.5}. We have
$$
\gathered
R_{s+1}(s+1,s+1)=g^t(s+1)\bmatrix \frac{s+z+\frac 12}{s+\frac 12}&0\\0& \frac{s+\frac 12}{s+z'+\frac 12}\endbmatrix
\Xi\, m_s^{-1}(s)\, \Xi^{-1} 
\left(I+\frac {B_s}{s+z+\frac 12}\right)\\ \times
\frac d{d\ze}\left(\left(I+\frac {B_s}{\ze+z-\frac 12}\right)^{-1}
\Xi\, m_s(\ze-1)\,\Xi^{-1} 
\bmatrix \frac{\ze -\frac 12}{\ze+z-\frac 12}&0\\0& \frac{\ze+z'-\frac 12}{\ze-\frac 12}\endbmatrix
\right)\Biggl|_{\ze=s+1}f(s+1).
\endgathered
\tag 7.6
$$

It is immediately seen that if the derivative falls on the last (diagonal) factor then the corresponding term vanishes. If the derivative falls on $m_s$ we obtain
$$
g^t(s+1)\bmatrix \frac{s+z+\frac 12}{s+\frac 12}&0\\0& \frac{s+\frac 12}{s+z'+\frac 12}\endbmatrix
\Xi\, m_s^{-1}(s) m_s'(s)\,\Xi^{-1} 
\bmatrix \frac{s+\frac 12}{s+z+\frac 12}&0\\0& \frac{s+z'+\frac 12}{s+\frac 12}\endbmatrix
f(s+1)
$$
which coincides with the right--hand side of \tht{7.4} because
$$
\Xi^{-1}\bmatrix \frac{s+\frac 12}{s+z+\frac 12}&0\\0& 
\frac{s+z'+\frac 12}{s+\frac 12}\endbmatrix
f(s+1)g^t(s+1)\bmatrix \frac{s+z+\frac 12}{s+\frac 12}&0\\0& \frac{s+\frac 12}{s+z'+\frac 12}\endbmatrix\Xi=f(s)g^t(s).
$$

Finally, if the derivative falls on $(I+B_s/(s+z+\frac 12))^{-1}$, we compute
$$
\multline
\frac d{d\ze}\left(I+\frac {B_s}{\ze+z-\frac 12}\right)^{-1}
=\left(I+\frac {B_s}{\ze+z-\frac 12}\right)^{-1}
\frac {B_s}{(\ze+z-\frac 12)^2}\left(I+\frac {B_s}{\ze+z-\frac 12}\right)^{-1}\\
=\frac 1{\ze+z-\frac 12}\left(I+\frac {B_s}{\ze+z-\frac 12}\right)^{-1}
\left(I-\left(I+\frac {B_s}{\ze+z-\frac 12}\right)^{-1}\right).
\endmultline
$$
Substituting into \tht{7.6} we obtain
$$
\gathered
R_{s+1}(s+1,s+1)=R_s(s,s)\\ -\frac 1{s+z+\tfrac 12}\,g^t(s)\,m_s^{-1}(s)
\,\Xi^{-1}
\left(I+\frac {B_s}{s+z+\frac 12}\right)^{-1}\Xi\, m_s(s)\,f(s).
\endgathered
\tag 7.7
$$

Taking determinants of both sides of \tht{6.3} we obtain 
$\det \left(I+{B_s}/({s+z+\frac 12})\right)=({s+z'+\frac 12})({s+z+\frac 12})\,,$
and hence
$$
\multline
\left(I+\frac {B_s}{\ze+z-\frac 12}\right)^{-1}=
\frac{s+z+\frac 12}{s+z'+\frac 12}\bmatrix
1+\frac{B_s^{22}}{s+z+\frac 12}&-\frac{B_s^{12}}{s+z+\frac 12}\\
-\frac{B_s^{21}}{s+z+\frac 12}&1+\frac{B_s^{11}}{s+z+\frac 12}
\endbmatrix\\ =\frac 1{s+z'+\frac 12}
\bmatrix 
s+z'+\frac 12-b_s&-\beta_s b_s\\ (b_s+z-z')/\beta_s& s+z+\frac 12+b_s
\endbmatrix.
\endmultline
$$
Substituting into \tht{7.7} we obtain
$$
\gathered
R_{s+1}(s+1,s+1)=R_s(s,s)-\frac 1{(s+z+\tfrac 12)(s+z'+\frac 12)}\\ \times \,g^t(s)\,m_s^{-1}(s)
\,\Xi^{-1}
\bmatrix 
s+z'+\frac 12-b_s&-\beta_s b_s\\ (b_s+z-z')/\beta_s& s+z+\frac 12+b_s
\endbmatrix
\Xi\, m_s(s)\,f(s).
\endgathered
\tag 7.8
$$

For $x\in\Z'_+$ denote
$$
m_x(x)=\bmatrix m_x^{11}&m_x^{12}\\m_x^{21}&m_x^{22}\endbmatrix.
$$
Since $\det m_x\equiv 1$ we have
$$
m_x^{-1}(x)=\bmatrix m_x^{22}&-m_x^{12}\\-m_x^{21}&m_x^{11}\endbmatrix.
$$
Substituting into \tht{7.8} we obtain
$$
\gathered
R_{s+1}(s+1,s+1)=R_s(s,s)-\frac {h_+^2(s)}{(s+z+\tfrac 12)(s+z'+\frac 12)}\\ \times \left(
\frac{\xi(b_s+z-z')}{\beta_s}\,{(m_s^{11})}^2+\frac{b_s\beta_s}{\xi}\, 
{(m_s^{21})}^2+(2b_s+z-z')m_s^{11}m_s^{21}\right).
\endgathered
\tag 7.9
$$

Similarly to the proof of Proposition 4.1, 
we now look at the residue of $m_{s+1}(\ze)$ at the point $\ze=s$. Since the jump matrix $-f(s)g^t(s)$ has zero first column, the residue itself also has zero first column. On the other hand, \tht{6.2} implies that
this residue equals $A_sm_s(s)$. Equating the (1,1) element of this matrix to zero and using the fact that $z+b_s\ne 0$ by the hypothesis of Proposition 6.3, we obtain 
$$
m_s^{11}=-\alpha_s\beta_s m_s^{21}.
\tag 7.10
$$ 
Substituting this relation into \tht{7.9} we arrive at \tht{7.2}.\qed
\enddemo

\demo{Proof of Lemma 7.4}
Let us look at the (1,1)--element of \tht{6.3} with $\ze=s+1$. We get
$$
\frac{s+\frac 12}{s+z+\frac 12}\,m_s^{11}=\left(1+\frac{b_s}{s+z+\frac 12}\right)m_{s+1}^{11}+
\frac{b_s\beta_s}{s+z+\frac 12} m_{s+1}^{21}.
$$
Using \tht{7.10} for $m_s^{11}$ and $m_{s+1}^{11}$ and simplifying we obtain 
$$
(s+\tfrac 12)\,
\alpha_s\beta_s\cdot\gamma_s=\left(b_s\beta_s-(b_s+s+z+\tfrac 12)\,\alpha_{s+1}\beta_{s+1}\right)\cdot \gamma_{s+1}.
$$
Substituting $\alpha_{s+1}$ and $\beta_{s+1}$ from \tht{6.10} and \tht{6.12} and simplifying further we arrive at \tht{7.3}.
\qed
\enddemo

\head 8. Initial conditions for dPV
\endhead
In this section we compute the initial conditions for the recurrences
\tht{6.10\,-12}. We will also show that the assumptions of Proposition 6.3 hold for generic values of parameters $(z,z',\xi)$.
\proclaim{Proposition 8.1} We have
$$
\gather
\alpha_{\frac 12}=-\frac{F(-z+1,\,-z';\,1;\,\xi)}{z'\xi\, F(-z+1,\,-z'+1;\,2;\,\xi)}\,,
\tag 8.1\\
b_{\frac 12}=-\frac{z\,F(-z+1,\,-z';\,1;\,\xi)}{F(-z,\,-z';\,1;\,\xi)}\,,
\tag 8.2
\\
\beta_{\frac 12}=-\frac{(zz'\xi)^\frac 12(1-\xi)^{z+z'}}{z\,F(-z+1,\,-z';\,1;\,\xi)}\,.
\tag 8.3
\endgather
$$
Here $F(a,\,b;\,c;\,u)$ is the Gauss hypergeometric function.
\endproclaim
\example{Remark 8.2} Using adjacency relations for the Gauss hypergeometric function, it is not hard to deduce from \tht{8.1}, \tht{8.2} the formula
$$
c_\frac 12=b_\frac 12+\frac{z'+1}{1-\xi\alpha_\frac 12}+z=\frac
{z'F(-z,\,-z'-1;\,1;\,\xi)F(-z+1,\,-z'+1;\,2;\,\xi)}
{F(-z,\,-z';\,1;\,\xi)F(-z+1,\,-z';\,2;\,\xi)}\,.
$$
(See Remark 6.4(4) for the definition of $\{c_s\}$.)
\endexample
\demo{Proof of Proposition 8.1} Similarly to Lemma 5.1, we have
$$
m_{\frac 12}(\ze)=\bmatrix 1&0\\-\sum_{x\in\Z'_-}
\frac{h_-^2(x)}{\ze-x}&1\endbmatrix.
$$
Similarly to the computation of $A_\frac 12$ in Proposition 5.2 we obtain
$$
A_\frac 12=-\frac{h_+^2(\tfrac 12)}{1+h_+^2(\tfrac 12)\sum_{x\in\Z'_-}\frac{h_-^2(x)}{(\frac 12-x)^2}}
\bmatrix \sum_{x\in\Z'_-}
\frac{h_-^2(x)}{\frac 12-x}&1 \\
-\left(\sum_{x\in\Z'_-}
\frac{h_-^2(x)}{\frac 12-x}\right)^2&-\sum_{x\in\Z'_-}
\frac{h_-^2(x)}{\frac 12-x}       
\endbmatrix.
\tag 8.4
$$
By \tht{6.2} we have
$$
m_{\frac 32}(\ze)=\left(I+\frac {A_{\frac 12}}{\ze-\frac 12}\right)m_{\frac 12}(\ze)=\left(I+\frac {A_{\frac 12}}{\ze-\frac 12}\right)
\bmatrix 1&0\\-\sum_{x\in\Z'_-}
\frac{h_-^2(x)}{\ze-x}&1\endbmatrix.
$$
Since $A_\frac 12$ is nilpotent we also have
$$
m_{\frac 32}^{-1}(\ze)=\bmatrix 1&0\\ \sum_{x\in\Z'_-}
\frac{h_-^2(x)}{\ze-x}&1\endbmatrix\left(I-\frac {A_{\frac 12}}{\ze-\frac 12}\right).
$$
Then \tht{6.3} gives
$$
\multline
\bmatrix 1&0\\ -\frac 1\xi\sum_{x\in\Z'_-}
\frac{h_-^2(x)}{\ze-x-1}&1\endbmatrix
\bmatrix \frac{\ze-\frac 12}{\ze+z-\frac 12}&0\\0& \frac{\ze+z'-\frac 12}{\ze-\frac 12} \endbmatrix
\bmatrix 1&0\\ \sum_{x\in\Z'_-}
\frac{h_-^2(x)}{\ze-x}&1\endbmatrix\left(I-\frac {A_{\frac 12}}{\ze-\frac 12}\right)\\=I+\frac{B_\frac 12}{\ze+z-\frac 12}\,.
\endmultline
$$
Taking the residue of both sides at $\ze=-z+\frac 12$ we obtain
$$
B_\frac 12=-\bmatrix 1&0\\ \frac 1\xi \sum_{x\in\Z'_-}
\frac{h_-^2(x)}{z+x+\frac 12}&0\endbmatrix \left(z\cdot I+A_\frac 12\right)\,.
$$
Thus, 
$$
\gather
b_\frac 12=B_\frac 12^{11}=-z-A_\frac 12^{11}=-z+\frac{h_+^2(\tfrac 12)\sum_{x\in\Z'_-}
\frac{h_-^2(x)}{\frac 12-x}}{1+h_+^2(\tfrac 12)\sum_{x\in\Z'_-}\frac{h_-^2(x)}{(\frac 12-x)^2}}\,,\\
b_\frac 12\beta_\frac 12=B_\frac 12^{12}=-A_\frac 12^{12}=\frac{h_+^2(\tfrac 12)}{1+h_+^2(\tfrac 12)\sum_{x\in\Z'_-}\frac{h_-^2(x)}{(\frac 12-x)^2}}\,.
\endgather
$$
Since
$(z+b_\frac 12)\alpha_\frac12\beta_\frac12=-A_\frac 12^{12}$ is equal to $b_\frac 12\beta_\frac 12$, we also obtain $\alpha_\frac 12=b_\frac 12/(z+b_\frac 12)$.

Now we recall the definition of $h_\pm$, see the beginning of \S6. We have 
$$
\gathered
h_+^2(\tfrac 12)=(zz'\xi)^\frac12(1-\xi)^{z+z'},\\
\sum_{x\in\Z'_-}
\frac{h_-^2(x)}{\frac 12-x}=(zz'\xi)^\frac12(1-\xi)^{-z-z'}\sum_{l=0}^\infty
\frac{(-z+1)_l(-z'+1)_l\xi^{l}}{l!^2(l+1)}\\=
(zz'\xi)^\frac12 (1-\xi)^{-z-z'}\,F(-z+1,\,-z'+1;\,2;\,\xi),
\\
1+h_+^2(\tfrac 12)\sum_{x\in\Z'_-}
\frac{h_-^2(x)}{(\frac 12-x)^2}=1+zz'\sum_{l=0}^\infty
\frac{(-z+1)_l(-z'+1)_l\xi^{l+1}}{(l+1)!^2}\\=
F(-z,\,-z';\,1;\,\xi).
\endgathered
$$
Hence,
$$
\gathered
b_\frac 12=-z+
\frac{zz'\xi\,F(-z+1,\,-z'+1;\,2;\,\xi)}{F(-z,\,-z';\,1;\,\xi)}=-\frac{z\,F(-z+1,\,-z';\,1;\,\xi)}{F(-z,\,-z';\,1;\,\xi)}\,,\\
\alpha_\frac12=\frac{b_\frac12}{b_\frac12+z}=
-\frac{\,F(-z+1,\,-z';\,1;\,\xi)}{z'\xi \,F(-z+1,\,-z'+1;\,2;\,\xi)}\,,\\
\beta_\frac12=\frac{B_\frac12^{12}}{b_\frac12}=
-\frac{(zz'\xi)^\frac 12(1-\xi)^{z+z'}}{z\,F(-z+1,\,-z';\,1;\,\xi)}\,.\qed
\endgathered
$$
\enddemo

The formulas of Propositions 6.3 and 8.1 allow us to extend the definition of the sequences $\{\alpha_s\}$, $\{b_s\}$ to arbitrary parameters $(z,z',\xi)\in \C\times\C\times (\C\setminus [1,+\infty))$ such that the denominators in \tht{6.10}, \tht{6.11}, \tht{8.1}, \tht{8.2} do not vanish. Now we will show that none of these denominators vanishes identically.   

According to Proposition 5.4, we can choose $\eta>0$ such that $w_s\ne 0$
for all $s\in\Z_+'$, where $w_s$ was defined in Corollary 3.3. Let us fix such an $\eta$ for the rest of this section, and let us also recall that for the sequence $\{v_s\}$ defined in Proposition 3.4, $v_s\ne 0,\pm1$, for all $s\in\Z'_+$. 

\proclaim{Proposition 8.3} Recurrence relations \tht{6.10}, \tht{6.11}, \tht{6.12} with initial conditions \tht{8.1}, \tht{8.2}, \tht{8.3} admit an asymptotic solution
of the form
$$
\gathered
|z|\to\infty, \quad |z'|\to\infty,\quad \xi\to 0,\qquad z,z'\in\C,\quad \xi\in(0,1),\\
z\,\xi^\frac 12=\eta+o(1),\quad  z'\xi^\frac 12=\eta+o(1),\\
\alpha_s=\xi^{-\frac 12}\left(\frac {v_{s-1}}{v_s}+o(1)\right),\quad b_s=-z-\eta\, v_{s-1}v_s+o(1),
\quad s=\tfrac 12,\tfrac 32,\dots\,,\\
\beta_\frac 12=-\xi^\frac 12(I_0^{-1}(2\eta)+o(1)),\quad
\beta_{s+1}=((1-v_s^2)^{-1}+o(1))\cdot {\beta_s},\quad s=\tfrac 12,\tfrac 32,\dots\,.
\endgathered
\tag 8.4
$$ 
\endproclaim
\example{Remark 8.4} It is not hard to see that in the limit \tht{8.4} the Lax pair \tht{6.2}, \tht{6.3} for dPV degenerates to the Lax pair \tht{3.4}, \tht{3.5} for dPII. This explains why solutions of dPII provide asymptotic solutions for dPV.
\endexample
\demo{Proof of Proposition 8.3} Using \tht{8.4} we obtain 
$$
\gathered
b_s+\dfrac{z'+s+\tfrac 12}{1-\xi\alpha_s}-\left(z'+s+\frac 12\right)=-z+O(1),
\\
b_s+\dfrac{z'+s+\tfrac 12}{1-\xi\alpha_s}-\left(z'-z+s+\frac 12\right)=
\frac{\eta v_{s-1}(1-v_s^2)}{v_s}+o(1),
\\
b_s+\dfrac{z'+s+\tfrac 12}{1-\xi\alpha_s}+z=z'+O(1),
\\
b_s+\dfrac{z'+s+\tfrac 12}{1-\xi\alpha_s}+z-z'=\frac{\eta v_{s-1}(1-v_s^2)}{v_s}+s+\tfrac 12+o(1).
\endgathered
$$
First three expressions are obviously nonzero while the last one is nonzero because \tht{3.13} implies that
$$
\frac{\eta v_{s-1}(1-v_s^2)}{v_s}+s+\tfrac 12=-\frac{\eta\,v_{s+1}(1-v_s^2)}{v_s}\,,
$$
and the sequence $\{v_s\}$ does not take values $0,\,\pm 1$.

Then \tht{6.10} turns into
$$
\frac{v_s}{v_{s+1}}=-\frac{\eta\, v_s(1-v_s^2)}{\eta v_{s-1}(1-v_s^2)+(s+\frac 12)v_s}+o(1)
$$
which holds by \tht{3.13}. Similarly, \tht{6.11} turns into
$$
-\eta\,v_s v_{s+1}=\eta\,v_{s-1}v_s-\frac{\eta v_{s-1}}{v_s}-s-\tfrac 12
-\frac{\eta v_{s+1}}{v_s}+o(1)
$$
which again holds by \tht{3.13}, and \tht{6.12} turns into the last relation in \tht{8.4}. 

The asymptotics for initial conditions is also immediate:
$$
\gather
\xi^\frac 12\alpha_{\frac 12}=-\frac{ F(-z+1,\,-z';\,1;\,\xi)}{z'\xi^\frac 12\, F(-z+1,\,-z'+1;\,2;\,\xi)}=-\frac {I_0(2\eta)}{I_1(2\eta)}+o(1)=\frac{v_{-\frac12}}{v_\frac 12}+o(1)
\\
z+b_{\frac 12}=\frac{z(F(-z,\,-z';\,1;\,\xi)-F(-z+1,\,-z';\,1;\,\xi))}{F(-z,\,-z';\,1;\,\xi)}\\ =\frac{zz'\xi\,F(-z+1,\,-z'+1;\,2;\,\xi)}{F(-z,\,-z';\,1;\,\xi)}
=\frac{\eta I_1(2\eta)}{I_0(2\eta)}+o(1)=-\eta\,v_{-\frac 12}v_\frac 12 +o(1),\\
\xi^{-\frac 12}\beta_\frac 12=-\frac{(zz'\xi)^\frac 12(1-\xi)^{z+z'}}{\xi^\frac12z\,F(-z+1,\,-z';\,1;\,\xi)}=-\frac 1{I_0(2\eta)}+o(1)\,.
\qed
\endgather
$$
\enddemo

Recall that if we define $D_s$ as a normalized Toeplitz determinant then it is an analytic functions in 
$(z,z',\xi)\in\C\times\C\times (\C\setminus [1,+\infty))$. 
Thus, now it makes sense to ask whether the relations \tht{7.1} are, in fact,  equalities of analytic functions. The answer is positive.

\proclaim{Theorem 8.5} Let $s\in \Z'_+$ and $D_s$ be the normalized  Toeplitz determinant defined in \S6. Define $\alpha_s$ and $b_s$ by the initial conditions \tht{8.1}, \tht{8.2} and recurrence relations \tht{6.10}, \tht{6.11}. Then for any $(z,z',\xi)\in\C\times\C\times (\C\setminus [1,+\infty))$ in the complement of the set of zeros of a nontrivial analytic function, the equality \tht{7.1} holds.
\endproclaim
\demo{Proof} Since both sides of \tht{7.1} are ratios of analytic functions, it suffices to prove that \tht{7.1} holds on some open set. 

Let us assume that $z'=\bar{z}\in \C\setminus \Z$ and $\xi\in (0,1)$. Fix $s\in\Z_+'$. Clearly, both sides of \tht{7.1} are ratios of analytic functions in $\Re z=(z+z')/2,\, \Im z=(z-z')/2,\, \xi$. According to Proposition 8.3, we can find some $\eta>0$ and small enough $\epsilon>0$ such that for 
$$
\xi\in(0,\epsilon),\quad \Re z\in(\eta\,\xi^{-\frac 12},\eta\,\xi^{-\frac 12}+1), \quad \Im z\in (1,2),
$$ 
the asymptotics \tht{8.4} ensures that neither numerators nor denominators in the formulas \tht{6.10}, \tht{6.11}, \tht{6.12} for the indices from $\frac 12$ to $s$ vanish. 
Thus, $\alpha_t$, $\beta_t$ for $t=\frac 12,\dots,s$ do not vanish as well.
Then Propositions 6.3 and 7.1 prove \tht{7.1} on this set of parameters. Complexifying $(\Re z,\, \Im z,\, \xi)$ proves \tht{7.1} on an open subset of $\C\times\C\times (\C\setminus [1,+\infty))$. \qed  
\enddemo

\head 9. Degeneration to continuous PII and PV 
\endhead

\subhead Discrete PII to continuous PII\endsubhead 
Although it does seem to be possible to degenerate the Lax pair \tht{3.4}, \tht{3.5} to a continuous limit, it is possible to find a scaling limit of the equation \tht{3.13} leading to the Painlev\'e II (ordinary differential)
equation, see e.g. \cite{ORGT}. Let us introduce a new real variable $t$ by 
$$
s=2\eta+\eta^\frac 13 t,\quad t=(s-2\eta)\,\eta^{-\frac 13}\,.
$$
Now let $\eta$ go to infinity, and assume that 
$v_s\approx (-1)^s\eta^{-\frac 13}v(t)$ as $\eta\to\infty$, with a smooth function $v(\,\cdot\,)$, and $s$ and $t$ related as above. Then we have
$$
\gathered
v_{s\pm1}=(-1)^{s+1}\eta^{- \frac13} \left(v(t)\pm\eta^{-\frac 13}v'(t)+\eta^{-\frac 23} v''(t)+O(\eta^{-1})\right),\\
\frac{(s+\frac 12)v_s}{\eta(v_s^2-1)}=(-1)^{s+1}
\left(2+\eta^{-\frac 23} t+\tfrac 12\eta^{-1}\right)
\eta^{-\frac 13}v(t)\left(1+\eta^{-\frac 23}v^2(t)+O(\eta^{-\frac 43})\right)\\
=(-1)^{s+1}\eta^{-\frac 13}\left(2v(t)+\eta^{-\frac 23}(tv(t)+2v^3(t))+O(\eta^{-1})\right).
\endgathered
$$
Substituting into \tht{3.13} and taking the limit $\eta\to\infty$ we get
$$
v''(t)=tv(t)+2v^3(t)
\tag 9.1
$$
which is a special case of Painlev\'e II equation. Since we also know, see \tht{5.6}, that $D_{s}D_{s+2}/D_{s+1}^2=1-v_s^2$, it is natural to assume that $D_s\approx D(t)$ for some smooth function $D(\,\cdot\,)$, and then we obtain 
$$
(\ln D(t))''=-v^2(t).
\tag 9.2
$$

As a matter of fact, the last formula is correct in the sense that there exists a solution $v(t)$ of \tht{9.1} such that $D_s=D(t)+o(1)$ as $\eta\to\infty$, and \tht{9.2} holds. This is a deep fact and it is the main result of \cite{BDJ1}.\footnote{The limit function $D(t)$ is known as the Tracy--Widom distribution in Random Matrix Theory, and it was obtained for the first time in \cite{TW1}.} For the history of this result, other proofs, generalizations, etc., we also refer to \cite{AD}, \cite{BDJ2}, \cite{BDR}, \cite{BOO}, \cite{BO3}, \cite{D2}, \cite{J2}, \cite{Ok}, \cite{W} and references therein.

\subhead Discrete PV to continuous PV
\endsubhead
The limit procedure considered in this section has a representation theoretic origin, see \cite{BO2, \S5}, \cite{BO3}, and also \cite{Bor2, \S8} for more details. 

We assume that $\xi\to 1$ and introduce a new complex variable $\omega$ and a new real variable $t$ by
$$
\omega=(1-\xi)\ze,\qquad t=(1-\xi)s.
$$
We will also redenote $m_s(\ze)$ as $m_t(\omega)$ and $A_s,\,B_s$
as $A(t),\, B(t)$ for $(\zeta,s)$ related to $(\omega,t)$ as above.
Let us assume that $m_t(\omega)$, $A(t)$, $B(t)$ all have smooth limits as $\xi\to 1$, $\zeta\to\infty$, $s\to+\infty$ in such a way that $\omega$ and $t$ converge to finite limits. Then the Lax pair equations \tht{6.2} and \tht{6.3} (in \tht{6.3} we use the right--hand side of \tht{6.2} instead of $m_{s+1}(\ze)$) tend to 
$$
\gathered
\frac{\partial m_t(\omega)}{\partial t}=\frac{A(t)}{\omega-t}\cdot m_t(\omega),\\
\frac{\partial m_t(\omega)}{\partial \omega}=
-\left(\frac{\sigma_3} 2 +\frac {B(t)}\omega+\frac{A(t)}{\omega-t}\right)m_t(\omega)+
m_t(\omega)\left(\frac 1\omega\bmatrix -z&0\\0&z'\endbmatrix +\frac {\sigma_3}2\right),
\endgathered
\tag 9.3
$$
where $\sigma_3=\bmatrix 1&0\\0&-1\endbmatrix$. Set 
$$
n_t(\omega)=m_t(\omega)\bmatrix \omega^{z}e^{-\frac \omega 2}&0\\ 0& \omega^{-z'}e^{\frac \omega 2}\endbmatrix.
$$
Then \tht{9.3} can be written in the form
$$
\gathered
\frac{\partial n_t(\omega)}{\partial t}=\frac{A(t)}{\omega-t}\cdot n_t(\omega),\\
\frac{\partial n_t(\omega)}{\partial \omega}=
-\left(\frac{\sigma_3} 2 +\frac {B(t)}\omega+\frac{A(t)}{\omega-t}\right)n_t(\omega),
\endgathered
\tag 9.4
$$
which is a  Lax pair for the Painlev\'e V equation, see \cite{JM, Appendix C}. The consistency relations \tht{6.7} tend to
$$
A'(t)=\frac{[A(t),\sigma_3]}2+\frac{[A(t),B(t)]}t\,,\qquad
B'(t)=\frac{[B(t),A(t)]}t\,,
\tag 9.5
$$
which are, quite naturally, the consistency relations for \tht{9.4}. The relations \tht{9.5} are usually called the {\it Schlesinger equations}. Now, if we parameterize the matrices $A(t)$ and $B(t)$ as
$$
A =(z+b )\bmatrix -1&-\alpha \beta \\ 1/(\alpha \beta )&1\endbmatrix,\quad
B =\bmatrix b &b \beta \\ (z'-z-b )/\beta &z'-z-b \endbmatrix,
$$
cf. \tht{6.8}, then the diagonal elements of \tht{9.5} give
$$
b'(t)=\frac{(z+b(t))\left(b(t)(1/\alpha(t)-\alpha(t))+(z'-z)\alpha(t)\right)}{t}\,.
\tag 9.6
$$
This is the limit of both \tht{6.19} and \tht{6.20}. Note that in the discrete case we derived the recurrence \tht{6.10}, \tht{6.11} using just the relations \tht{6.19}, \tht{6.20}. In the continuous limit \tht{6.19} becomes equivalent to \tht{6.20}, and we need additional arguments. 

Instead of deriving Painlev\'e V in the usual way by a more careful examination of \tht{9.5}, we will take a shortcut and use the limit of the relation \tht{6.11}. (Note that the limit of \tht{6.10} in the first order approximation is trivial.) We obtain
$$
2b(t)=\frac{t(\alpha'(t)+\alpha(t))}{(1-\alpha(t))^2}+\frac{z-z'}{1-\alpha(t)}-2z+z'.
\tag 9.7
$$

Substituting into \tht{9.6} yields
$$
\gathered
\alpha''(t)=\left(\frac 1{2\alpha(t)}+\frac 1{\alpha(t)-1}\right)(\alpha'(t))^2-\frac{\alpha'(t)}t+\frac{(z-z'-1)\alpha(t)}t\\+\frac{(\alpha(t)-1)^2}{2t}\left((z')^2\alpha(t)-{z^2}/\alpha(t)\right)-\frac 12\,\frac{\alpha(t)(\alpha(t)+1)}{\alpha(t)-1}
\endgathered
\tag 9.8
$$
which is the Painlev\'e V equation in the standard form. 

Now if we assume that $D_s\approx D(t)$ with a smooth $D(\,\cdot\,)$, then the relation \tht{7.1} becomes trivial in the first order approximation, and the second asymptotic term gives a rather cumbersome expression for
$(\ln D(t))'''/(\ln D(t))''$ in terms of $\alpha(t)$, $b(t)$, and their first derivatives. \footnote{We use the fact that for any smooth function $f(x)$, 
$$
\left(\frac{f(x+\epsilon)}{f(x)}
-\frac{f(x+2\epsilon)}{f(x+\epsilon)}\right)/\left(\frac{f(x+2\epsilon)}{f(x+\epsilon)}-\frac{f(x+3\epsilon)}{f(x+2\epsilon)}\right)=1-\frac{(\ln f(x))'''}{(\ln(f(x))''}\,\epsilon+O(\epsilon^2).
$$
}

In fact, it is known that under the limit transition described above, the Fredholm determinant $D_s$ of the hypergeometric kernel tends to the Fredholm determinant $D(t)=\det(1-K_t)$ where $K_t$ is the {\it Whittaker kernel} restricted to $(t,+\infty)$, see \cite{BO1}, \cite{Bor1} for the definition of the Whittaker kernel, and \cite{BO2}, \cite{BO3} for the limit transition. Moreover, as was shown in \cite{BD, \S8}, $D(t)$ is the {\it isomonodromy $\tau$-function} of the Schlesinger equations \tht{9.5}. This means that there exist solutions $\alpha(t)$, $b(t)$ of \tht{9.7}, \tht{9.8} such that  $(\ln D(t))'$ is expressed in terms of $\alpha(t)$ and $\beta(t)$ in the following simple way:
$$
t(\ln D(t))'=\operatorname{Tr}(AB)+\tfrac t2\operatorname{Tr} (A\sigma_3)=(b+z)(z'-z+(b+z-z')\alpha+b/\alpha-t).
\tag 9.9
$$
Using \tht{9.5} one also computes
$$
(t(\ln D(t))')'=\tfrac 12\operatorname{Tr} (A\sigma_3)=-(b+z).
$$
It remains unclear whether there exist a discrete analog of either of these simple formulas. It is also worth noting that the function $\sigma(t)=t(\ln D(t))'$ itself satisfies a second order differential equation
$$
(t\si'')^2=(2(\si')^2-t\si'+\si+
(z+z')\si')^2-
4(\si')^2(\si'+z)(\si'+z').
\tag 9.10
$$ 
This result was first proved by C.~Tracy \cite{T} using the method of \cite{TW2}; it also follows from \tht{9.7}--\tht{9.9}. The equation \tht{9.10} is the so--called $\sigma$-form of the Painlev\'e V equation. It is also not clear if there exists a discrete analog of
\tht{9.10}.

\vskip 1 true cm

School of Mathematics, Institute for Advanced
Study, Einstein Drive, Princeton NJ 08540, U.S.A.

E-mail address:
{\tt borodine\@math.upenn.edu}

\newpage
$$\epsffile{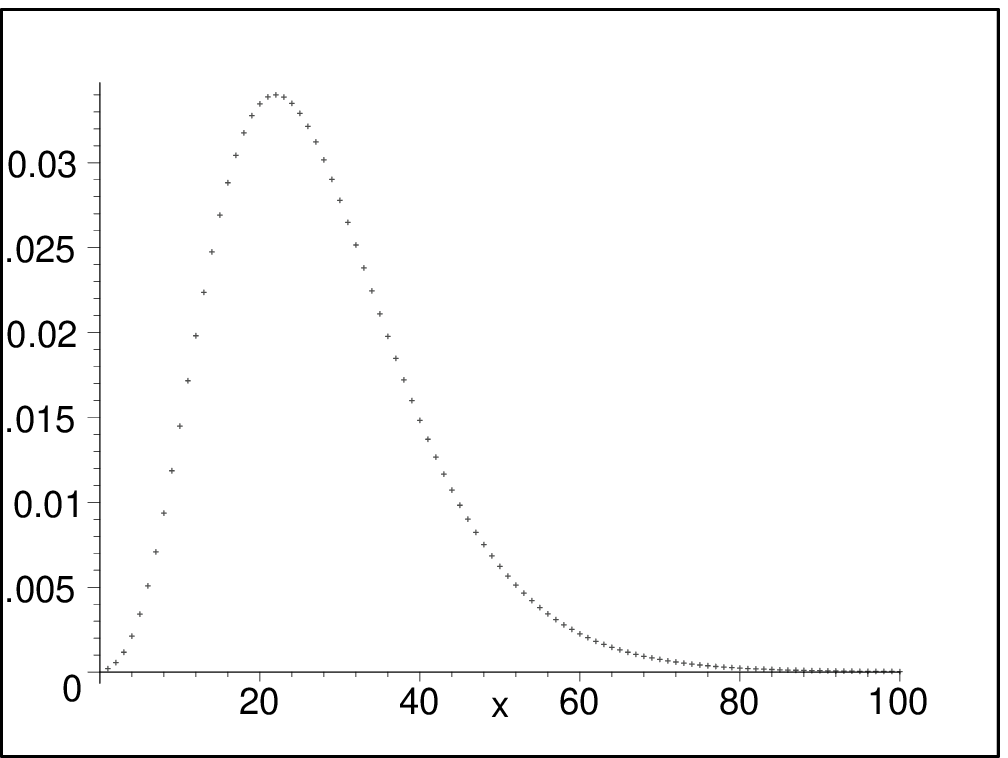}$$
\centerline{The plot of  $q_{x+1}^{(z,z',\xi)}-q_{x}^{(z,z',\xi)}$ for  $z=z'=2.5$, $\xi=0.85$, see  Introduction.}

\enddocument